\documentclass{aa}
\usepackage{graphicx}
\usepackage{txfonts}
\begin{document}
   \title{An ISOCAM survey through gravitationally lensing galaxy
          clusters\thanks{Based on observatios with ISO, an ESA
          project with instruments funded by ESA Member States
          (especially the PI countries: France, Germany, the
          Netherlands and the United Kingdom) with the participation
          of ISAS and NASA}}
   \subtitle{II. The properties of infrared galaxies in the A2218 field}

   \author{A. Biviano\inst{1}
   \and L. Metcalfe\inst{2,3}
   \and B. McBreen\inst{4}
   \and B. Altieri\inst{2}
   \and D. Coia\inst{4}
   \and M.F. Kessler\inst{5}
   \and J-P. Kneib\inst{6}
   \and K. Leech\inst{3}
   \and K. Okumura\inst{7}
   \and S. Ott\inst{5}
   \and R. Perez-Martinez\inst{2,3}
   \and C. Sanchez-Fernandez\inst{2,3}
   \and B. Schulz\inst{8}
          }

\offprints{A. Biviano}

\institute{INAF/Osservatorio Astronomico di Trieste, via G.B. Tiepolo 11, 
34131, Trieste, Italy, \email{biviano@ts.astro.it}
\and XMM-Newton Science Operations Centre, European Space Agency,
Villafranca del Castillo, P.O. Box 50727, 28080 Madrid, Spain
\and ISO Data Centre, European Space Agency, Villafranca del Castillo, 
P.O. Box 50727, 28080 Madrid, Spain
\and Physics Department, University College Dublin, Stillorgan Road, 
Dublin 4, Ireland
\and Science Operations and Data Systems Division of ESA, ESTEC, 
Keplerlaan 1, 2200 AG Noordwijk, The Netherlands
\and Observatoire Midi-Pyr\'en\'ees, 14 Av. E. Belin, 31400 Toulouse,
France
\and DSM/DAPNIA Service d'Astrophysicque, CEA-Saclay, B\^at. 709,
Orme des Merisiers, 91191 Gif-sur-Yvette Cedex, France
\and Infrared Processing and Analysis Center, California Institute of
Technology, Pasadena, CA 91125, USA
             }

   \date{Received ; accepted }

   \abstract{We have observed the cluster Abell 2218 ($z=0.175$)
   with ISOCAM on board the {\em Infrared Space Observatory} using two
   filters, LW2 and LW3, with reference wavelengths of 6.7 and 14.3
   $\mu$m, respectively. We detected 76 sources down to 54 and 121
   $\mu$Jy (50\% completeness levels) at 6.7 and 14.3 $\mu$m,
   respectively. All these sources have visible optical
   counterparts. We have gathered optical and near-infrared magnitudes
   for 60 of the 67 non-stellar optical counterparts to the ISOCAM
   sources, as well as redshifts for 43 of them. We have obtained
   acceptable and well constrained fits to the observed spectral
   energy distributions (SEDs) of 41 of these sources, using the
   ``GRASIL'' models of Silva et al. (1998), and have determined their
   total infrared luminosities ($L_{IR}$'s) and star formation rates
   (SFRs).

   The SEDs of 20 (out of 27) ISOCAM cluster members are best fit by
   models with negligible ongoing star formation, and no
   major episode of star formation in the last $\sim 1$ Gyr. Their
   SEDs resemble those of 5--10 Gyr old early-type galaxies. A slightly
   higher, but still very mild, star-formation activity is found among
   the remaining cluster sources, which are mostly spirals. The
   median IR luminosity of the 27 ISOCAM cluster sources is $L_{IR}
   = 6 \times 10^8 \, L_{\odot}.$ The ISOCAM-selected cluster
   galaxies have indistinguishable velocity and spatial distributions
   from those of the other cluster galaxies, and do not contribute
   significantly to the Butcher-Oemler effect. If A2218 is undergoing
   a merger, as suggested by some optical and X-ray analyses, then
   this merger does not seem to affect the mid-infrared properties of its
   galaxies.

   The SEDs of most ISOCAM-selected field sources are best fit by
   models with moderate ongoing star formation, with a significant
   fraction of their stellar mass formed in the last $\sim 1$ Gyr.
   Their SEDs resemble those of massive star-forming spirals or
   starburst galaxies, observed close to the maximum of their star
   formation activity, but not necessarily during the short-lived
   starburst event. The median redshift of these field galaxies is
   $z \simeq 0.6$. Their $L_{IR}$'s span almost two orders of
   magnitudes, from $\sim 10^{10} \, L_{\odot}$ to $\sim 10^{12} \,
   L_{\odot}$, with a median of $1.2 \times 10^{11}$ (eight of the 14
   field sources are LIRGs). The SFRs of these 14 ISOCAM-selected
   field sources range from 2 to 125 $M_{\odot} \, {\mbox yr}^{-1}$,
   with a median value of 22 $M_{\odot} \, {\mbox yr}^{-1}$.

   We compare our findings with those obtained in other ISOCAM cluster
   and field surveys.

   \keywords{Galaxies: clusters: general -- Galaxies: clusters: individual
    (Abell 2218) -- Infrared: galaxies} }

   \maketitle
%

\section{Introduction}\label{s-intro}
Clusters of galaxies are not a very common galaxy environment, but
they are crucial for our understanding of galaxy formation and
evolution. In a hierarchical cosmological scenario, the first galaxies
to form are those located in high density peaks, which then evolve to
become galaxy clusters (e.g., Coles et al. \cite{cole99}). Hence, on
average, cluster galaxies are believed to be older than field galaxies
(e.g., Kauffmann \& Charlot \cite{kauf98}).  When a field galaxy is
accreted by a cluster, it is likely to speed up its evolution rate, as
a consequence of the physical processes which are switched on in the
new, rather extreme, environmental conditions, such as ram-pressure,
tidal stripping, `harassment', and galaxy-galaxy collisions (see,
e.g. Abraham et al. \cite{abra96}; Acreman et al. \cite{acre03};
Gnedin \cite{gned03}). While the long-term effect of these processes
is to transform a star-forming galaxy into a quiescent one (or to
disrupt it entirely, in the case of dwarfs), it is yet unclear how
this transformation occurs, and, in particular, if the affected galaxy
undergoes a starburst phase during the transformation process (see,
e.g., Poggianti \cite{pogg03}; Okamoto \& Nagashima
\cite{okam03}).

A direct observational evidence of the evolution of the properties of
cluster galaxies with redshift $z$, is the Butcher-Oemler (BO,
hereafter) effect , i.e. the excess of blue galaxies in distant
clusters relative to nearby ones (see, e.g., Butcher \& Oemler
\cite{butc84}; Margoniner et al. \cite{marg01}; Pimbblet
\cite{pimb03}). The effect is partly, yet not entirely, due to field
contamination (see, however, Andreon et al. \cite{andr04} who argue
that the BO-effect could be due to observational biases).  The
BO-effect is based on photometric data. The spectroscopic analogue of
the BO-effect is the increase of the fraction of `k+a' galaxies
(originally called `E+A' by Dressler \& Gunn \cite{dres83}) with $z$.
The spectra of k+a galaxies are characterized by strong Balmer lines
in absorption and the absence of (strong) emission lines (see
e.g. Dressler et al. \cite{dres99}; Ellingson et
al. \cite{elli01}). These spectroscopic features are usually
interpreted as the signatures of a post-starburst phase (see,
e.g. Poggianti \cite{pogg03}). Another direct evidence of the
evolution of cluster galaxies is the change in the morphological mix
of clusters with $z$ (Dressler et al. \cite{dres97}; Fasano et al.
\cite{fasa00}); clusters at $0.2 \leq z \leq 0.6$ have a significantly
lower fraction of S0s (or early-type galaxies, see Fabricant et
al. \cite{fabr00}) than nearby clusters. This suggests that a
transformation of spiral galaxies into S0s took place over the last
$\sim 4$ Gyr.

It is yet unclear how the BO-effect is related to the change in the
morphological fractions of cluster galaxies. The blue BO galaxies
seem to be mostly disk galaxies with disturbed morphologies (Lavery \&
Henry \cite{lave94}).  They could be spirals seen before they
transform into S0s. If the transformation occurs after the gas content
of these galaxies is used in one or several starbursts, an excess
population of post-starburst galaxies is naturally expected at an
intermediate phase of the transformation process. Since starbursts
usually occur in very dusty environments, it is important to explore
these processes at infrared (IR) wavelengths, where the stellar
radiation is re-emitted, in the presence of dust (see, e.g. Poggianti
\cite{pogg03}).  In particular, it was shown by Poggianti et
al. (\cite{pogg99}) that a particular spectral class of cluster
galaxies, the `e(a)' galaxies (characterized by strong Balmer lines in
absorption and emission lines of moderate intensity), are most easily
interpreted as dust-hidden starbursts. If the e(a) galaxies are the
progenitors of the more common k+a spectral-type galaxies, then it is
likely that also k+a galaxies are dust-rich.

Up to now, the available information on the IR properties of cluster
galaxies is rather limited. Mid-IR (MIR, hereafter) fluxes have been
published for members of the nearby clusters Virgo, Coma, and A1367
(Boselli et al. \cite{bose97}, \cite{bose98}; Quillen et
al. \cite{quil99}; Contursi et al. \cite{cont01}), as well as for the
medium-distant clusters A370 (Soucail et al. \cite{souc99}; Metcalfe
et al. \cite{metc03}, hereafter Paper~I), A1689 (Fadda et
al. \cite{fadd00}; Duc et al. \cite{duc02}), A1732 (Pierre et
al. \cite{pier96}), A2218 (Barvainis et al. \cite{barv99}; Paper~I),
A2219 (Barvainis et al. \cite{barv99}), A2390 (L\'emonon et
al. \cite{lemo98}; Altieri et al. \cite{alti99}; Paper~I), and
Cl0024+1654 (Coia et al. \cite{coia04}, C04 hereafter). These analyses
have shown that, on average, optically-selected normal early-type
galaxies have their MIR emission dominated by the Rayleigh-Jeans tail
of the cold stellar component, while late-type galaxies have their MIR
emission dominated by the thermal emission from dust (Boselli et
al. \cite{bose98}).  In the Coma cluster, the k+a galaxies were only
found to have enhanced 12~$\mu$m emission when also optical
emission-lines were present (Quillen et al. \cite{quil99}). Since
early-type galaxies are the dominant cluster population, one does not
expect to find a large number of strong-IR emitters among cluster
galaxies at wavelengths where the stellar photospheric emission
becomes negligible. Nevertheless, in C04 we report the detection of a
large number of very bright IR-galaxies in the distant cluster
Cl0024+1654, an indication that the IR properties of cluster members
can vary with redshift, and among different clusters.

Outside clusters, there is now ample evidence for a significant
evolution of the population of MIR-detected field galaxies (Altieri et
al. \cite{alti99}; Aussel et al. \cite{auss99}; Elbaz et
al. \cite{elba99}, \cite{elba02}, \cite{elba03};
Paper~I; Sato et al. \cite{sato03}, Serjeant et
al. \cite{serj00}). These MIR field galaxies have a redshift
distribution peaked around $z \sim 0.7$.  Their MIR-emission is mostly
powered by starbursts (rather than AGNs), with typical star formation
rates (SFRs) of $\sim 50 \, M_{\odot} \, {\mbox yr}^{-1}$. The
integrated source counts of the MIR galaxy population exceed by an
order of magnitude the predictions of no-evolution models based on
the local MIR luminosity function (Franceschini et al. \cite{fran01};
Paper~I).  Models that explain the high MIR source counts require
either a strong evolution of the whole luminosity function or a strong
evolution in both luminosity and density of a sub-population of
starburst galaxies (Chary \& Elbaz \cite{char01}; Franceschini et
al. \cite{fran01}; Xu \cite{xu00}).  The integrated emission from the
field MIR sources detected by ISOCAM accounts for most of the cosmic
infrared background detected by Puget et al. (\cite{puge96}; see Elbaz
et al. \cite{elba02}).

In this paper we analyse the properties of IR-selected galaxies in the
A2218 cluster, located at a mean redshift $z=0.175.$
We base our analysis on the data obtained in our
gravitational-lensing deep-survey programme, which was part of the
``Central Programme'' of the ISO mission. These data were used in
Paper~I, in conjunction with ISOCAM data for other clusters, to derive
deep source counts at 6.7 and 14.3 $\mu$m. The counts we derived in
Paper~I confirm and extend earlier findings of an excess by a factor
of ten in the population of 14.3 $\mu$m emitters with respect to
no-evolution models (Altieri et al.  \cite{alti99}; Aussel et
al. \cite{auss99}; Elbaz et al. \cite{elba99}), and they agree with
the counts of 6.7 $\mu$m sources reported by Sato et
al. (\cite{sato03}) and Aussel et al. (\cite{auss99}).  We complement
the ISOCAM MIR data with photometric and spectroscopic data at optical
and near-IR (NIR) wavelengths, taken from the literature. We fit the
optical--MIR spectral energy distributions (SEDs) of the ISOCAM
sources with the GRASIL models of Silva et al. (\cite{silv98}, S98
hereafter) to constrain the spectral types, total IR luminosities, and
SFRs of the cluster and background sources.

In \S~\ref{s-data} we describe our data-set. In \S~\ref{s-sed} we
describe the model fits to the observed SEDs of the ISOCAM
extragalactic sources, and in \S~\ref{s-sfr} we describe how we
determine the IR luminosities and SFRs of these sources. In the
two following sections we describe the results of our analysis,
separately for cluster sources (\S~\ref{s-cl}) and field sources
(\S~\ref{s-fd}).  We address the issue of the BO-effect in the IR in
\S~\ref{s-bo}. In \S~\ref{s-disc} we discuss our findings, and compare
them with previous results from the literature. In \S~\ref{s-conc} we
summarize our results and provide our conclusions.  Some aspects of
the SED fitting procedure are detailed in Appendix~\ref{a-sedfit}. We
discuss a few individual ISOCAM sources and their SEDs in
Appendix~\ref{a-sources}.

\begin{figure*}
\centering
\resizebox{\hsize}{!}{\includegraphics{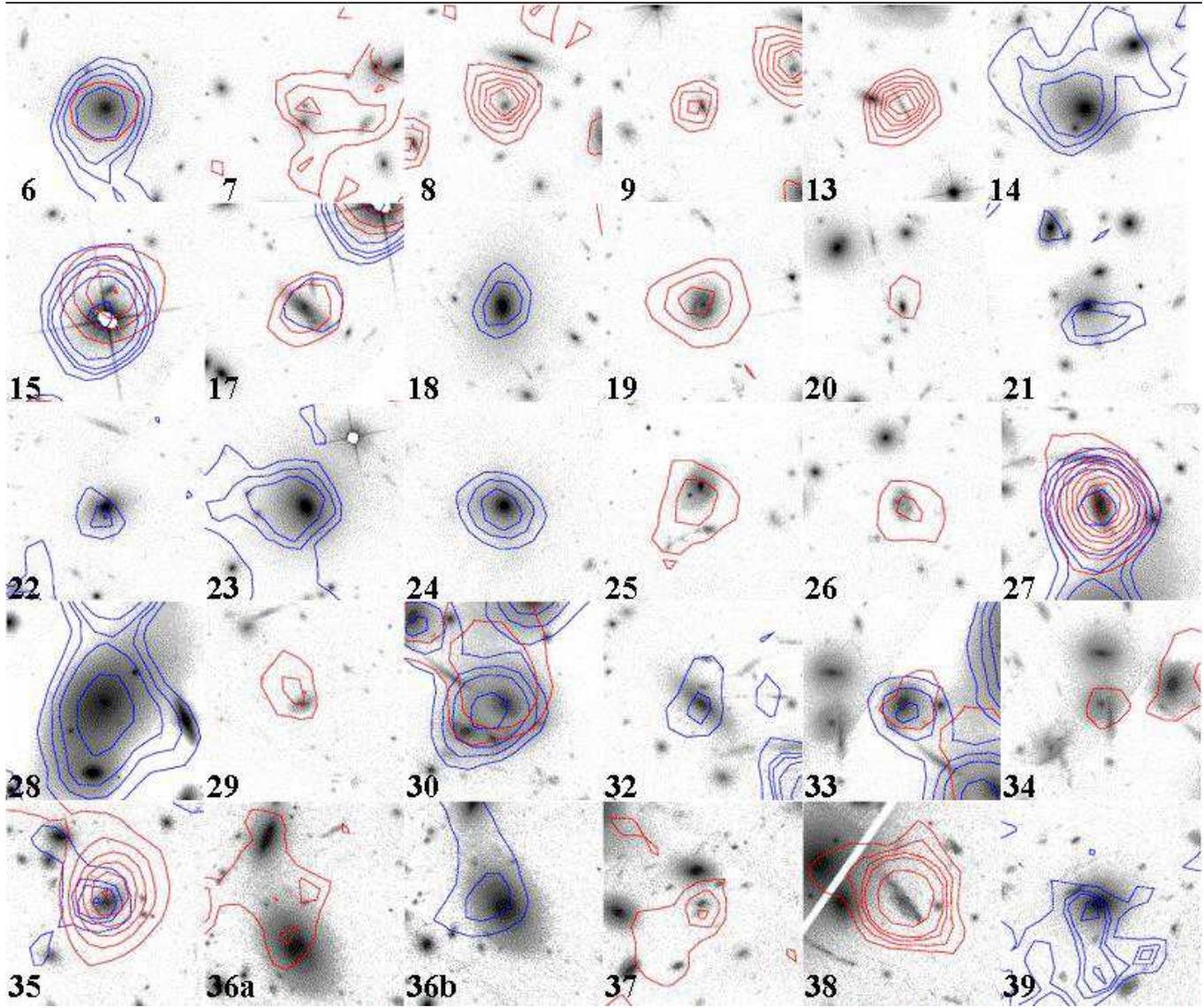}}
\caption{MIR isocontours of ISOCAM sources,
at 6.7 and 14.3 $\mu$m (in blue and red,
respectively) overlaid on an optical HST image in
the F814W band.}
\label{f-imag1}
\end{figure*}
\addtocounter{figure}{-1}

\begin{figure*}
\centering
\resizebox{\hsize}{!}{\includegraphics{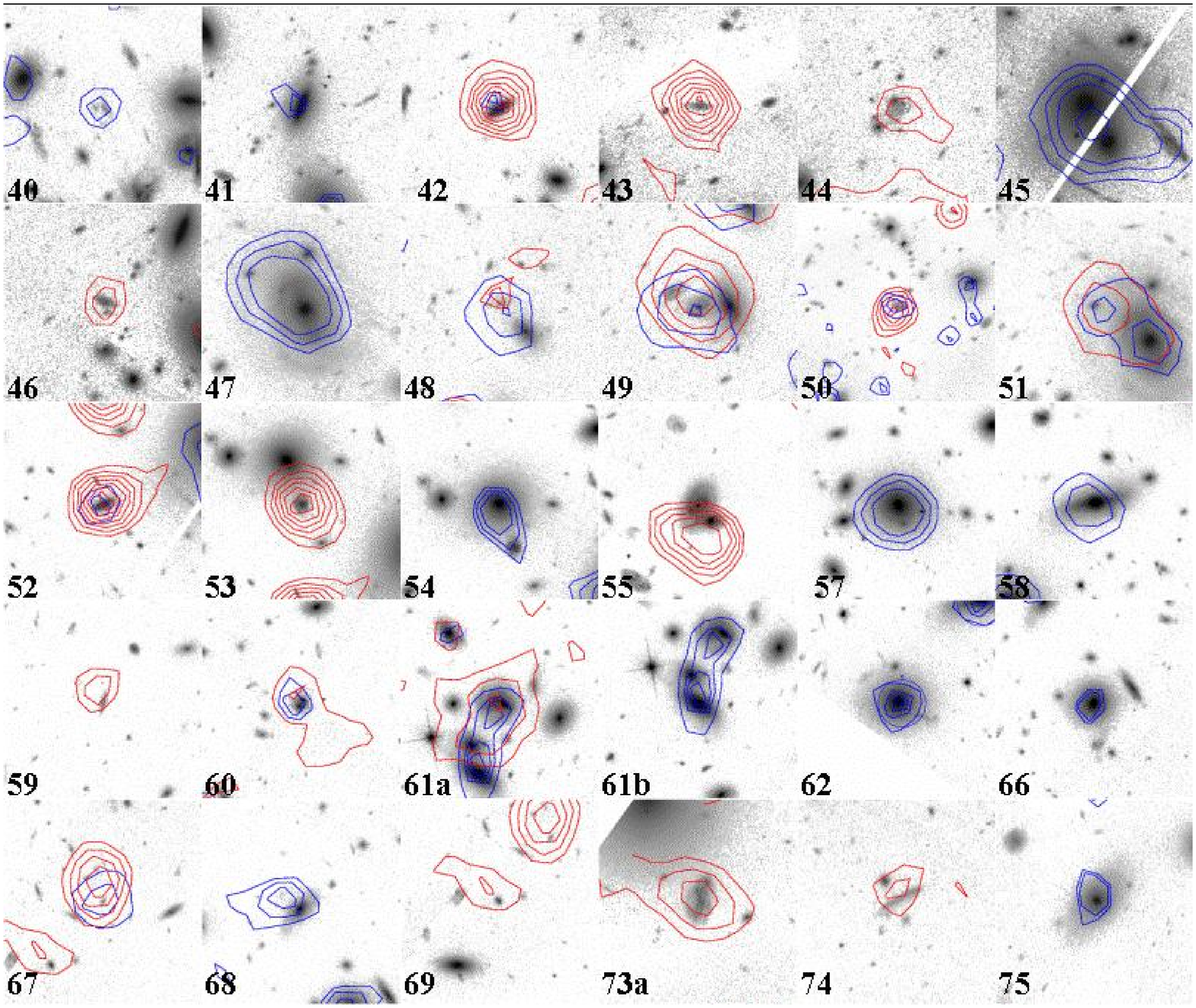}}
\caption{Continued.}
\end{figure*}

In this paper we use $H_0=70$ km~s$^{-1}$~Mpc$^{-1}$, $\Omega_0=0.3$
and $\Omega_{\Lambda}=0.7$ (see, e.g., Melchiorri \& \"Odman
\cite{melc03}). In this cosmology, the cluster luminosity distance is
845~Mpc, 1 arcmin corresponds to 178~kpc, and the age of the Universe
at the cluster redshift is 11.3 Gyr.

\section{The Data}\label{s-data}
The Abell 2218 field was observed with the LW2 and LW3 filters of
ISOCAM (Cesarksy et al. \cite{cesa96}), with reference wavelengths of
6.7 and 14.3 $\mu$m, respectively, in the context of a
gravitational-lensing deep survey programme, part of the ``Central
Programme'' of the ISO mission. This programme is described in detail
in Paper~I. The total observing time was 6.2 hours per filter.  The
observations were done in raster mode with the $3''$ per-pixel
field-of-view, and a raster step size of $16''$, covering a total area
of 20.5 arcmin$^2$, i.e. $\simeq 0.8 \times 0.8$ Mpc$^2$ at the
cluster distance.

For a detailed description of the ISOCAM data reduction and source
identification and reliability, we refer the reader to Paper~I. The
only modification with respect to Paper~I concerns an improved
photometry for the three extended sources 28, 30, and 45 (we use the
ISOCAM source identification numbers of Paper~I). As was pointed out
in Paper~I in the cases of sources 28 and 45, the fluxes of these
three sources as listed in that paper are underestimates, because their
MIR photometry was derived under the point-source assumption, but
these sources are extended.  The total flux of these sources is
between 1.5 (source 45) and 2 (sources 28 and 30) times higher than
quoted in Paper~I. Since these three sources are cluster members,
their flux corrections have no effect on the field-source counts
reported in Paper~I.

In total, 76 sources were detected in the A2218 field, with an average
level of significance (i.e. the ratio of the source signal to the
1-$\sigma$ local noise floor) of 9 and 12 in the LW2 and the LW3 band,
respectively. Four of these sources were also detected by Barvainis et
al. (\cite{barv99}); their flux density estimates are in agreement
with ours, within the errors. Among our 76 MIR sources, 18 were
detected in both bands, 30 in the LW2 band only, and 28 in the
LW3 band only. Source fluxes at 6.7 and 14.3 $\mu$m are in the ranges
23--934 $\mu$Jy, and 90--919 $\mu$Jy, respectively. Background sources
have their fluxes amplified by the cluster gravitational
lensing. After correction for lensing amplification (see Paper~I), the
faintest source detected at 6.7 (14.3) $\mu$m has an intrinsic flux of
5 (18) $\mu$Jy.

We looked for optical and near-infrared (NIR) counterparts for the 76
ISOCAM sources, in the catalogues of Le Borgne et al. (\cite{lebo92},
hereafter LPS), Smail et al. (\cite{smai01}, hereafter S01), and
Ziegler et al. (\cite{zieg01}, hereafter Z01). The catalogue of LPS
contains $B, g, r, i,$ and $z$ magnitudes for 729 objects in the
central $4 \times 4$ arcmin$^2$ of A2218. LPS also provide redshifts
for 66 objects in the same field.  The catalogue of S01 contains
photometry in the F450W, F606W, and F814W filters of the {\em Hubble
Space Telescope} -- HST herafter --, and in the $K_s$ filter, for 81
galaxies in the central $2.5 \times 2.5$ arcmin$^2$ of A2218.  S01
also provide morphologies for most of their sources, and redshifts for
34 of them. Finally, the catalogue of Z01 contains $U, B, V, I$
magnitudes and redshifts for 48 early-type cluster members in the
central $2.5 \times 2.5$ arcmin$^2$, as well as morphologies for 19
galaxies.  

We estimated the morphologies of another 17 optical
counterparts from an HST image in the F814W band. The HST mosaic of
Abell 2218 was observed with WFPC2 in the F606W filter under the GO
programme 7343 (PI: Squires). Most of the observations were conducted
in March and May 1999, and the remaining in May 2000.  It comprises 24
pointings observed in the CVZ (continue viewing zone) during 2
orbits. At each position 12 exposures of 700 sec were obtained
totalizing 8.4 ksec.  The data were retrieved from the ST-ECF/ESO
archive.  The data reduction was done using standard IRAF and STSDAS
scripts and was combined into one mosaic using Swarp
(www.terapix.fr/swarp).  More details on these data can be found in
Hudelot et al. (\cite{hude04}). For those ISOCAM sources that lie in the field
covered by this HST image, we show in Fig.~\ref{f-imag1} their MIR
isocontours at 6.7 and 14.3 $\mu$m (in blue and red, respectively)
overlaid on the optical image (sources with stellar counterparts --
see below -- are not shown).

We search for optical/NIR counterparts within a circle of $12''$
radius around each ISOCAM source, in order to match (roughly) the size
of the ISOCAM PSF diameter at 14.3 $\mu$m (Okumura
\cite{okum98}). Following the method of Flores et al. (\cite{flor99}),
we determine quantitative likelihoods of all possible
counterparts. These likelihood estimates are based on the distance
between the ISOCAM source and its possible counterpart, and on the
magnitude of the possible counterpart (see Flores et al. \cite{flor99}
for details). The mean (median) distance between the ISOCAM sources
and their most likely counterparts is $1.4''$ ($1.2''$), and all
sources, except two, have distances $<4''$ (sources 5 and 60 have
distances of $\sim 7''$ from their most likely optical
counterparts). Visual inspection of the ISOCAM LW2 and LW3 maps
overlaid on optical images (see Fig.~\ref{f-imag1} and Figs. 6 and 7
in Paper~I) confirms the identification of the counterparts obtained
by the maximum likelihood procedure. The likelihoods of the assigned
optical/NIR counterparts of our ISOCAM sources are very high. Cases
where confusion might play a relevant r\^ole are detailed in
Appendix~\ref{a-sources}.

Since the ISOCAM flux densities are total flux densities, we must
convert the apparent magnitudes of the optical/NIR counterparts into
total magnitudes.  S01 provide total magnitudes in the $K_s$-band, and
$B_{450}-I_{814}$, $V_{606}-I_{814}$, $I_{814}-K_s$ aperture colours,
that we use to derive approximate total $B_{450}$, $V_{606}$, and
$I_{814}$ magnitudes (assuming negligible colour gradients within the
galaxies).  Similarly, we determine approximate total $U$, $B$, and
$V$ magnitudes, using the $I$-band total magnitudes and $U-I$, $B-I$,
$V-I$ aperture colours in Z01. We then convert Z01's magnitudes into
the photometric system of S01, using the mean magnitude difference (in
bands of similar central wavelength) of objects in common to the two
data-sets.  Finally, we transform LPS' isophotal magnitudes into
total, by using the following relations between LPS' and S01's
magnitudes, determined for objects in common to the two data-sets:
$B_{S01}^{total}=B_{LPS}-0.60$, $I_{S01}^{total}=i_{LPS}-1.04$,
$V_{S01}^{total}=0.5 \times (r_{LPS}+g_{LPS})-0.50$. Clearly, the
magnitudes obtained using these empirical relations have larger errors
than the original ones, because of the scatter in the relations ($\sim
0.2$ magnitudes), and these larger errors are taken into account in
our analysis.

All the optical and NIR magnitudes are corrected for galactic
extinction, and then converted into flux densities using the relations
of Fukugita et al. (\cite{fuku95}) and Tokunaga (\cite{toku00}).

Nine of the 76 ISOCAM sources are identified with stars. These are
sources no. 3, 4, 11, 12, 15b, 16, 56, 71, and 72. Seven of our ISOCAM
sources are outside the surveyed areas of the LPS, S01, or Z01
catalogues. These are sources no. 19, 36a, 43, 48, 50, 55, and
73b. The lack of photometry in optical/NIR bands precludes a thorough
analysis of their properties. These 16 sources are not considered in
this paper. We are thus left with a sample of 60 extragalactic ISOCAM
sources with optical counterparts and available optical and/or NIR
magnitudes. 

\section{The Spectral Energy Distributions}\label{s-sed}
We construct Spectral Energy Distributions (SEDs) for the 60
non-stellar ISOCAM sources with available optical/NIR photometry.
When either the 6.7 or the 14.3 $\mu$m flux density is not available,
an upper limit is used instead. This limit is equal to the 50\%
completeness limit for the region where the ISOCAM source is located
(see Table~3 in Paper~I), or twice that value for the extended sources
28, 30, and 45. With this definition, MIR upper limits approximately
correspond to 4~$\sigma$ limits. SED models that violate this limit
are rejected in the fitting procedure.

\begin{table*}
\centering
\caption[]{Results of the model SED fitting}
\label{t-sed}
\begin{tabular}{lllllllcc}
\hline
ISOCAM & Optical         & morphology  & Best-fit     & Nsfr(0.1) & Nsfr(1.1) & Redshift    & Membership &
Quality \\
source id. & counterpart  & (reference) & model  &     & & (reference) &             &
of the fit \\
\hline
ISO\_A2218\_02  & L664            & --          & Sa5  & 0.4               & 0.5  & 0.53 (Paper~I)        & f & P \\
ISO\_A2218\_05  & L621            & Sa          & Sa5  & 0.4 (0.1--0.8)    & 0.1 (0.1--3.4)  & $0.175 \pm 0.15$ & c & G \\   
ISO\_A2218\_06  & L630            & E/S0        & Em5  & 0.0 ($\leq 0.3$)  & 0.0 ($\leq 0.6$)  & 0.1802 (LPS)     & c & G \\   
ISO\_A2218\_08  & L587            & Sc          & Mps2 & 1.4 (1.1--3.0)    & 1.3 (1.3--1.4)   & 0.68 (Paper~I)        & f & G \\
ISO\_A2218\_09  & L565            & SBa         & Sa2  & 1.1 (0.4--1.8)    & 1.2 (0.5--2.4)   & 0.68 (Paper~I)        & f & G \\  
ISO\_A2218\_10  & L557            & --          & Sb5  & 1.1 ($\leq 1.8$)  & 1.2 ($\leq 3.4$) & $0.00 \pm 0.37$  & - & U \\   
ISO\_A2218\_13  & L537            & S           & M5   & 3.0 (1.1--3.0)    & 1.3 (1.3--1.4)   & $0.90 \pm 0.16$  & f & G \\
ISO\_A2218\_14  & L535/Z2076      & E           & Es2  & 0.0 ($\leq 0.3$)  & 0.00 ($\leq 0.6$)  & 0.1827 (LPS)     & c & G \\
ISO\_A2218\_17  & L484            & Sa          & Sb10 & 0.6 (0.4--1.1)    & 0.6 (0.5--3.4)   & $0.10 \pm 0.08$  & c & G \\
ISO\_A2218\_18  & L482/S4013      & E/S0        & Em5  & 0.0               & 0.0              & 0.1778 (LPS)     & c & G \\
ISO\_A2218\_20  & L467/S159       & S0 (S01)    & Aps2 & 0.5 ($\leq 1.1$)  & 1.6 (0.1--2.4)  & 0.476  (E98)     & f & G \\   
ISO\_A2218\_21  & L436/S137/Z1516 & SB0/a (S01) & Em5  & 0.0 ($\leq 0.3$)  & 0.0 ($\leq 0.6$)  & 0.1638 (S01)     & c & G \\   
ISO\_A2218\_22  & L430/S337/Z2604 & E           & Em5  & 0.0 ($\leq 0.3$)  & 0.0 ($\leq 0.6$)  & 0.1800 (LPS)     & c & G \\   
ISO\_A2218\_23  & Z1976           & E           & Em2  & 0.0 ($\leq 0.3$)  & 0.1 ($\leq 0.6$)  & 0.1686 (Z01)     & c & G \\   
ISO\_A2218\_24  & L419/Z1142      & E           & Es2  & 0.0 ($\leq 0.3$)  & 0.0 ($\leq 0.6$)  & 0.1641 (LPS)     & c & G \\   
ISO\_A2218\_25  & L409            & SB0/a       & Em2  & 0.0 ($\leq 0.3$)  & 0.1 ($\leq 0.6$)   & 0.1741 (LPS)     & c & G \\   
ISO\_A2218\_26  & L411/S103       & Sdm (S01)   & Sa5  & 0.4 ($\leq 1.8$)  & 0.5 ($\leq 3.4$)  & $0.175 \pm 0.20$ & c & G \\   
ISO\_A2218\_27  & L395/S333       & Sc (S01)    & Sa5  & 0.4 (0.4--0.8)    & 0.5 (0.5--3.4)   & 0.1032 (LPS)     & f & G \\   
ISO\_A2218\_28  & L391/S301       & cD (S01)    & Em10 & 0.0               & 0.0                & 0.1720 (LPS)     & c & P \\   
ISO\_A2218\_29  & L381            & SB0/a       & Sc2  & 1.8 (0.4--1.8)    & 1.4 (0.5--2.4)   & 0.521  (E98)     & f & G \\
ISO\_A2218\_30  & L373/S280/Z1552 & SB0/a (S01) & Sa10 & 0.1 ($\leq 0.1$)  & 0.1 ($\leq 0.1$)  & 0.1776 (LPS)     & c & G \\   
ISO\_A2218\_32  & L348/Z2270      & Sa (Z01)    & Es10 & 0.0 ($\leq 0.3$)  & 0.0 ($\leq 0.6$)  & 0.1676 (Z01)     & c & G \\   
ISO\_A2218\_33  & L341/S351/Z1662 & E (S01)     & Em10 & 0.0               & 0.0               & 0.1637 (LPS)     & c & G \\   
ISO\_A2218\_34  & L323            & S           & Sa10 & 0.1 ($\leq 0.3$) & 0.1 ($\leq 0.6$)  & 0.179  (E98)     & c & G \\   
ISO\_A2218\_35  & L317/S420       & Scd (S01)   & Mps5 & 1.1               & 1.4                & 0.474  (E98)     & f & P \\   
ISO\_A2218\_36b & L296/Z1888      & Sab (Z01)   & Em2  & 0.0 ($\leq 0.3$)  & 0.1 ($\leq 0.6$)  & 0.1760 (LPS)     & c & G \\   
ISO\_A2218\_38  & L289            & S/Irr       & Mps5 & 1.1 (1.1--3.0)    & 1.4 (1.3--1.4)   & 1.033  (LPS)     & f & G  \\   
ISO\_A2218\_39  & Z849            & E           & El5  & 0.0 ($\leq 0.3$)  & 0.0 ($\leq 0.6$)  & 0.1652 (Z01)     & c & G  \\   
ISO\_A2218\_40  & L287            & Irr         & Sb2  & 1.6 (0.4--3.0)    & 1.4 (0.5--2.4)   & 0.702  (E98)     & f & G  \\   
ISO\_A2218\_42  & L275            & Sa          & Sb2  & 1.6 (1.1--1.8)    & 1.4 (1.2--1.5)   & 0.45   (Paper~I)      & f & G \\   
ISO\_A2218\_44  & L262            & Irr         & Sc2  & 1.8               & 1.4                & 0.596  (E98)     & f & G \\   
ISO\_A2218\_45  & L244/S307/Z1437 & S0 (S01)    & Em10 & 0.0               & 0.0                & 0.1646 (LPS)     & c & G \\        
ISO\_A2218\_47  & L235/Z1175      & SB0/a (Z01) & Em10 & 0.0 ($\leq 0.3$)  & 0.0 ($\leq 0.6$)  & 0.1765 (LPS)     & c & G \\   
ISO\_A2218\_52  & L206/S323       & Sc (S01)    & Sb5  & 1.1 (0.4--1.1)    & 1.2 (0.5--1.4)   & 0.55   (Paper~I)      & f & G \\   
ISO\_A2218\_53  & L205/S368       & Sc (S01)    & Sc5  & 1.6               & 1.5                & 0.693 (E98)      & f & P  \\
ISO\_A2218\_54  & L196/S401/Z1466 & E (S01)     & Em5  & 0.0 ($\leq 0.3$)  & 0.0 ($\leq 0.6$)  & 0.1798 (LPS)     & c & G  \\   
ISO\_A2218\_57  & L149/S1057      & E (S01)     & Em10 & 0.0 ($\leq 0.3$)  & 0.0 ($\leq 0.6$)  & 0.1753 (LPS)     & c & G  \\   
ISO\_A2218\_58  & L148/S428/Z1343 & S0 (S01)    & Em10 & 0.0               & 0.0                & 0.1830 (LPS)     & c & G  \\   
ISO\_A2218\_59  & L145/S646       & Sdm (S01)   & Mps5 & 1.1 (0.4--1.1)    & 1.4 (0.5--1.4)   & 0.628  (E98)     & f & G  \\   
ISO\_A2218\_60  & L116/S4010      & E (S01)     & Sa5  & 0.4 ($\leq 1.1$)  & 0.5 ($\leq 3.4$)  & 0.2913 (LPS)     & f & G  \\   
ISO\_A2218\_61a & L119/S638       & S0 (S01)    & Sa10 & 0.1 ($\leq 0.8$)  & 0.1 ($\leq 2.4$)  & $0.175 \pm 0.12$ & c & G  \\   
ISO\_A2218\_61b & L113/S633/Z2702 & S0 (S01)    & Es10 & 0.0               & 0.0                & 0.1738 (LPS)     & c & G  \\   
ISO\_A2218\_62  & L118/Z1293      & E (Z01)     & Em5  & 0.0 ($\leq 0.3$)  & 0.0 ($\leq 0.6$)  & 0.1758 (Z01)     & c & G  \\   
ISO\_A2218\_66  & L77/S4007       & SB0 (S01)   & Es5  & 0.0 ($\leq 0.3$)  & 0.0 ($\leq 0.6$)  & 0.1681 (LPS)     & c & G  \\        
ISO\_A2218\_67  & L75/S4004       & --          & Es2  & 0.0               & 0.0                & 1.5              & - & P \\  
ISO\_A2218\_68  & L62/S4009       & Sa (S01)    & Sa5  & 0.4 ($\leq 1.1$)  & 0.5 ($\leq 2.4$)  & 0.42   (Paper~I)      & f & G  \\   
ISO\_A2218\_70  & L58             & --          & Sa10 & 0.1               & 0.1                & 0.1703 (LPS)     & c & G  \\   
ISO\_A2218\_73a & L38/S4015       & Sc (S01)    & M10  & 2.8 ($\leq 11.9$) & 1.0 ($\leq 3.4$)  & $0.30 \pm 0.22$ & - & U  \\
\hline
\end{tabular}
\end{table*}

In order to fit the observed SEDs of our ISOCAM sources we consider a
battery of models computed with the GRASIL code of S98.  This code
computes the spectral evolution of stellar systems by taking into
account the effects of dust.  In particular, it takes into account
several environments with different dust properties and distributions,
such as the AGB envelopes, the diffuse interstellar medium, and the
molecular clouds. GRASIL models consider different kinds of dust
particles, including the very small grains and the polycyclic aromatic
hydrocarbons that are mainly responsible for the MIR emission
detected by ISOCAM.  GRASIL models have been shown to provide
excellent fits to the observed SEDs of nearby galaxies, in both
quiescent and starburst phases, with infrared data obtained with ISO
(S98; Mann et al. \cite{mann02}; C04). Full details of the GRASIL
model are given in S98.

\begin{figure}
\centering
\resizebox{\hsize}{!}{\includegraphics{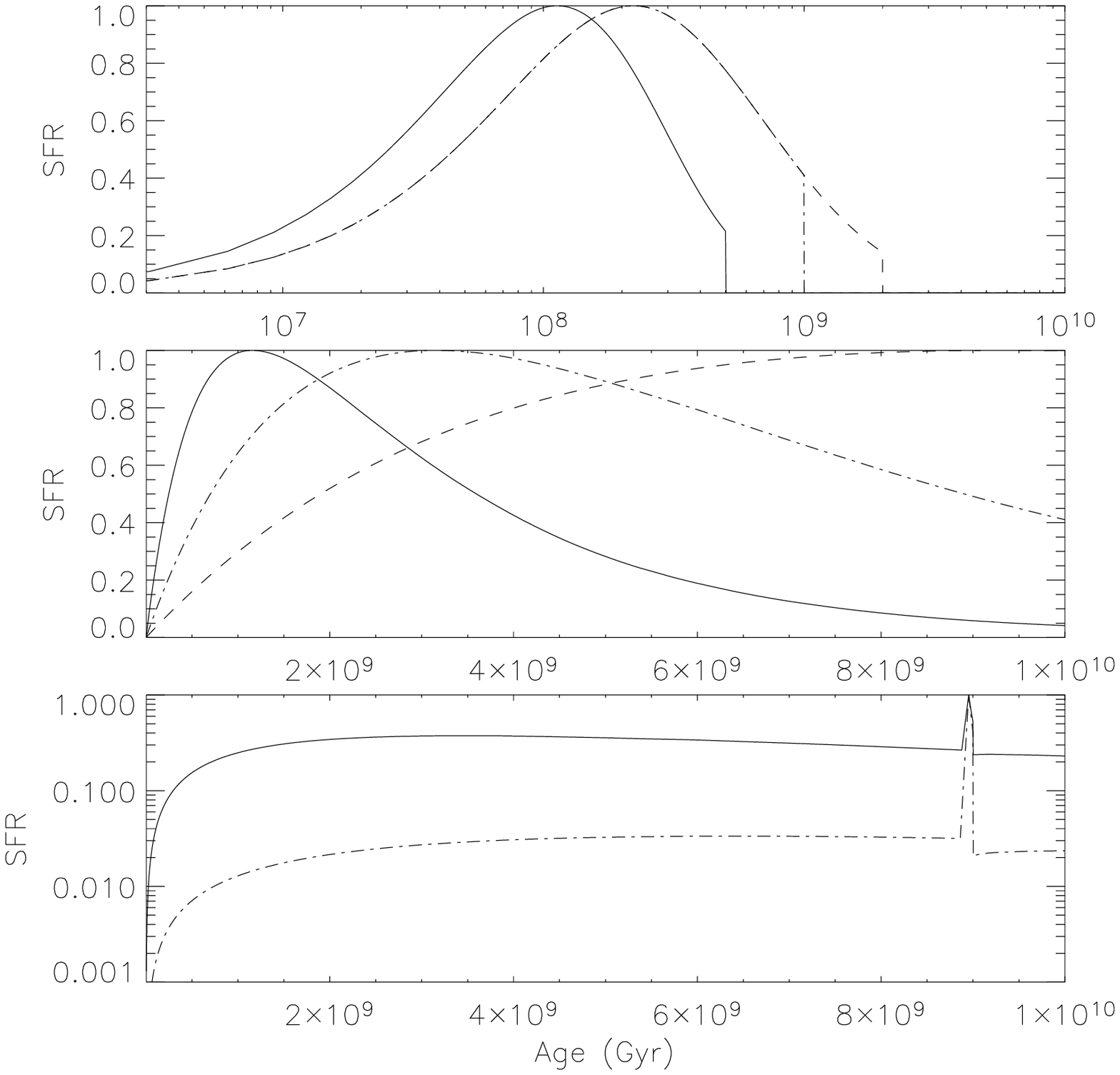}}
\caption{The star formation histories of the GRASIL models used in
this paper. Time since galaxy formation is on the x-axis, and SFR
(normalised to the maximum value) is on the y-axis. Note the
logarithmic x- and y-scale in the top and bottom panel, respectively.
Top panel: Es (solid line), Em (dash-dotted line), and El (dashed
line) models.  Middle panel: Sa (solid line), Sb (dash-dotted line),
and Sc (dashed line) models.  Bottom panel: Mps (solid line), and Aps
(dash-dotted line) models.  Models M and A are like models Mps and
Aps, respectively, except that they are seen at the time of the
starburst (the spike in the Figure) rather than 1 Gyr after it.  See
the text for a detailed description of all models.}
\label{f-sfh}
\end{figure}

The available photometric data for our MIR-selected sources do not
allow an accurate modelling of each individual galaxy SED.  Rather, we
perform the comparison of the observed SEDs with a limited set of
models meant to be representative of four broad classes of spectral
types. Mann et al. (\cite{mann02}) have considered five
model SEDs, while we have considered 20 model SEDs in C04, i.e. 10
models each seen at two different ages.  Here we consider 30 models,
i.e. the same 10 models of C04, each seen at three different
ages. Some of these models were taken directly from the GRASIL web
site\footnote{http://web.pd.astro.it/granato/grasil/grasil.html \\
or: http://adlibitum.oat.ts.astro.it/silva/grasil/grasil.html}, some were
kindly provided by L.~Silva (private comm.), and some were built by
ourselves by running the GRASIL code (also freely available on the
web).

\begin{figure}
\centering
\resizebox{\hsize}{!}{\includegraphics{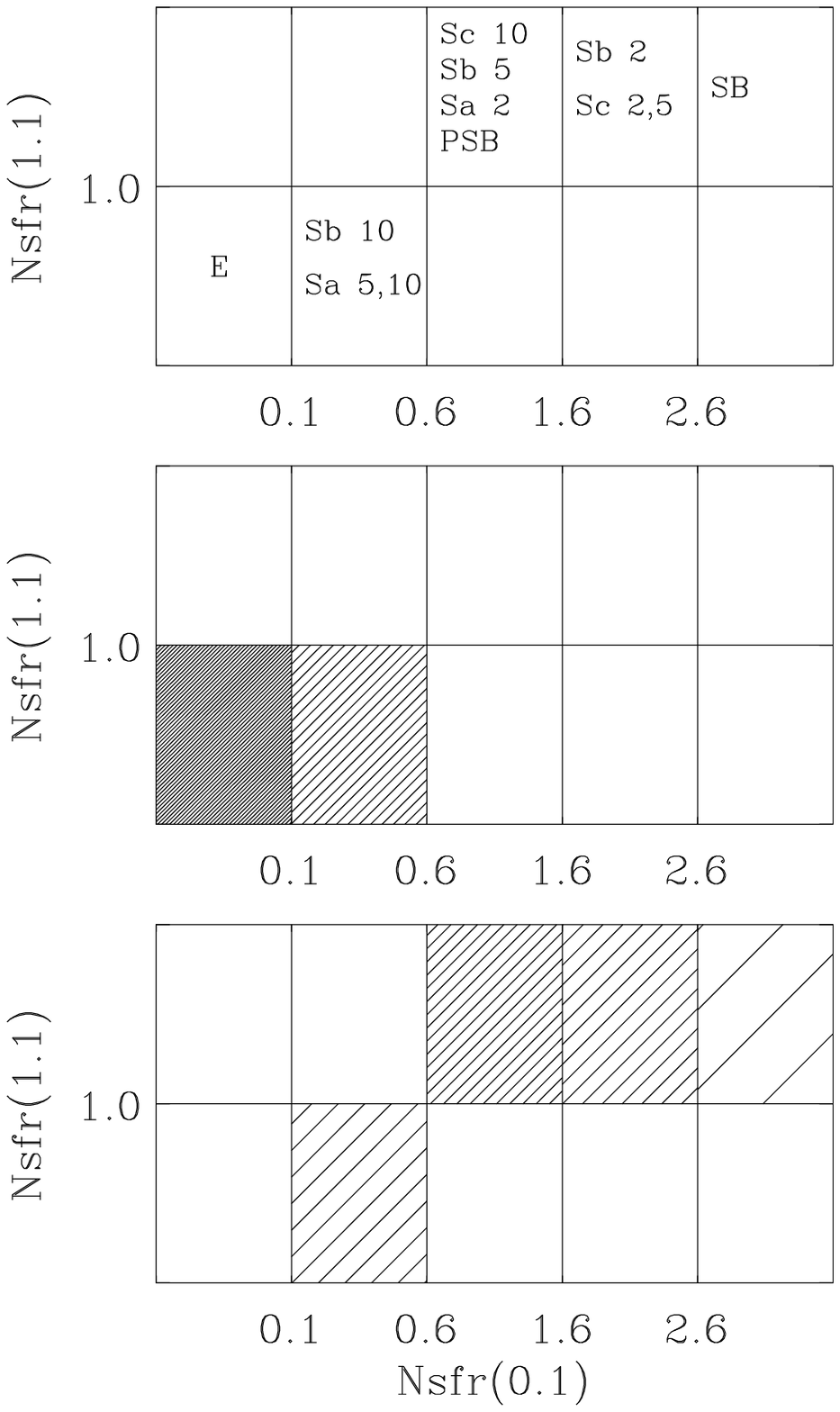}}
\caption{Top panel: The SFR averaged over the last 1.1 Gyr vs. the SFR
averaged over the last 0.1 Gyr, both normalised by the SFR averaged
over the entire SFH, for the different GRASIL models, grouped in five
classes. Middle panel: the distribution of the best-fit models for the
ISOCAM cluster sources in the Nsfr(1.1) vs. Nsfr(0.1) plane. The
shading in each box is proportional to the number of best-fit
models. Bottom panel: same as the middle panel, but for ISOCAM field
sources.}
\label{f-fstar}
\end{figure}

The models we consider are the following:
\begin{itemize}
\item[{\bf E:}] three models of early-type galaxies, characterized by
an initial burst of star formation lasting 0.5, 1.0, or 2.0 Gyr, and
by passive evolution thereafter. Depending on the duration of the
initial burst, we label these models `Es' (short), `Em' (medium), and
`El' (long). These models have been used in Granato et
al. (\cite{gran01}) and reproduce the range of SEDs of ellipticals
and S0s in the nearby universe.
\item[{\bf S:}] three models of disk galaxies (spirals), characterized
by different values of: i) the gas infall timescales, and ii) the
efficiency term in the Schmidt-type law (see \S~2.1 in S98). The star
formation histories (SFHs, herafter) of these models gently
increases with time up to a maximum, and then gently decreases.  The
maximum SFR occurs at $\sim 1$, $\sim 3$, and $\sim 9$ Gyr after
galaxy formation, for the three models here considered, which we label
`Sa', `Sb', and `Sc', respectively, because their SEDs reproduce those
of nearby spirals of the Sa, Sb, and Sc types.
\item[{\bf SB:}] two starburst models that provide good fits to the observed
SEDs of the moderate starburst galaxy M82, and of the strong starburst
galaxy Arp220. We refer to these two starburst models as `M' and `A',
respectively.  The starburst is characterized by an e-folding time of
0.05 Gyr, and involves $\sim 0.01$ and $\sim 0.1$ of the total mass of
the galaxy, for the M and A model, respectively.
\item[{\bf PSB:}] the two starburst models mentioned above (`M' and `A'),
but observed 1 Gyr after the starburst event; we label these 'post-starburst'
models `Mps' and `Aps'.
\end{itemize} 

The SB models we consider here are rather extreme examples of the
starburst phenomenon. We considered including NGC6090 as another
example of starburst galaxy, of intermediate IR luminosity between
Arp220 and M82.  However, after suitable normalisation, the SED of M82
is not different enough from that of NGC6090 to justify including an
additional model in our analysis, given the relatively large error bars on 
our (rather faint) MIR fluxes.

The SFHs of the 10 models are shown in Figure~\ref{f-sfh}. Three
SEDs were built for each of these ten models, by stopping their star
formation histories at ages of 2, 5, and 10 Gyr. The SB and PSB model
SEDs are obtained from the same models, stopping their star formation
histories at the time of the starburst event, and, respectively, 1 Gyr
later.

Since these GRASIL models were produced in order to fit the observed
SEDs of nearby galaxies, it is not clear what the physical meaning of,
e.g., a 2 Gyr Sa model is. Such a model might, or might not, resemble
a real Sa of that age, depending on the way an Sa galaxy forms and
evolves.  It is therefore more meaningful to parametrize these models
in terms of their current and recent SFR's. More specifically, we
define the normalized star formation rates Nsfr(0.1) and Nsfr(1.1) as
the SFR of the model, averaged over the
last 0.1 and, respectively, 1.1 Gyr, divided by the SFR averaged over
the entire galaxy SFH. These two quantities measure the {\em shape} of
the model SFH, rather than the {\em intensity} of the star formation
process. Using Nsfr(0.1) and Nsfr(1.1) we can group our 30 models into
five classes (see Figure~\ref{f-fstar}, top panel). Values of
$\mbox{Nsfr(1.1)} < 1$ characterize galaxies that have experienced
their main star formation activity more than 1.1 Gyr ago.  Among
these, those with $\mbox{Nsfr(0.1)} > 0.1$ have some residual star
formation activity left.  When $\mbox{Nsfr(1.1)} \geq 1$, we
distinguish three cases, depending on the value of Nsfr(0.1). When
Nsfr(0.1) is very high, the galaxy is undergoing a
starburst. Intermediate values of Nsfr(0.1) characterize galaxies near
a maximum of star forming activity in their SFH (see
Figure~\ref{f-sfh}). Finally, when Nsfr(0.1) is comparable or even
lower than Nsfr(1.1), the galaxy might have recently suffered a
starburst, but the starburst phase is now over. These are the models
that we call post-starbursts (PSB), although they still retain
substantial amounts of star formation activity, as the starburst event
has used up only a fraction of the available gas. Note that the range
of Nsfr(0.1) and Nsfr(1.1) values that characterizes the PSB galaxies
is also shared by other models, namely those of normal spirals
observed shortly after the peak of their SFH.

We compare the observed and model SEDs with a standard $\chi^2$
procedure (see Appendix~\ref{a-sedfit} for a detailed description of
some technical aspects of our SED-fitting procedure). When the
redshift of the optical counterpart to the ISOCAM source is known,
there are two free parameters in the fit, the model SED and its
normalization. When the redshift is unknown, it is taken as an
additional free parameter. We only accept the best-fit solution if the
range of acceptable values for the source redshift is sufficiently
narrow (see Appendix~\ref{a-sedfit}).

In order to constrain the fit to a galaxy SED and estimate its
probability, the number of free parameters plus one must be less than
the number of photometric data-points. This is true for 48 of the
original 60 sources. For 40 of them, the redshift is known. 

Note that, currently, no AGN component is included in the GRASIL
models.  We do not expect to find many cases of AGN-dominated emission
among MIR selected sources. Several studies (e.g., Franceschini et
al. \cite{fran01}; Rowan-Robinson \cite{rowa01}) have shown that most
ISOCAM sources have their MIR emission dominated by dust-reprocessed
stellar radiation, especially at the lower flux levels. Elbaz et
al. (\cite{elba02}) have found that only $12 \pm 5$\% of the ISOCAM sources
in the Hubble Deep Field North have their MIR flux dominated by dust
re-processed AGN radiation. Among the 48 ISOCAM sources for which we
try a SED-model fitting, we can therefore expect $6 \pm 2$ in which
the MIR emission is dominated by dust-reprocessed radiation from an
obscured AGN. This number could be different if the relative number
of AGNs among galaxies depends on the environment.

The results of our SED-fitting analysis are summarized in
Table~\ref{t-sed}, along with the optical counterparts and their
morphologies when available. In the columns of Table~\ref{t-sed} the
following information is listed:
\begin{enumerate}
\item ISOCAM source identification;
\item identification of the optical/NIR counterpart in the LPS, S01,
and Z01 catalogues (L: LPS, S: S01, Z: Z01); 
\item morphology of the optical counterpart, and reference
(S01, Z01; if no reference is given, the morphology estimate is ours);
\item best-fit model, followed by a number indicating the age
of the model in Gyr;
\item Nsfr(0.1), i.e. the SFR of the best-fit model, averaged over the
last 0.1 Gyr, divided by the SFR averaged over the entire
SFH of the model. The 95~\% confidence level (c.l. hereafter) range (in brackets)
is also given, when applicable, i.e. when other models provide an
acceptable ($\chi^2$-probability of the fit $\geq 0.05$) fit to the
observed SED, and when the Nsfr(0.1) values of these models
are different from the best-fit one;
\item same as Col.~5, but for Nsfr(1.1) (the SFR averaged over the
last 1.1 Gyr, divided by the SFR averaged over the entire
SFH), in lieu of Nsfr(0.1);
\item spectroscopic redshift, if available, and reference (LPS, S01,
Z01, Paper~I, and Ebbels et al. \cite{ebbe98},
hereafter E98), or photometric redshift obtained through the SED
fitting procedure, with its uncertainty (rms of the redshift values
found among all the acceptable fits, see Appendix~\ref{a-sedfit}), if
available;
\item cluster/field membership (labelled c/f, respectively),
based on the spectroscopic or photometric
redshift estimates of Col.~6;
\item quality of the fit (G: good, $\chi^2$-probability of the fit 
$\geq 0.05$; P: poor, $\chi^2$-probability of the fit 
$< 0.05$; U: unconstrained, $\chi^2$-probability of the fit 
$\geq 0.05$, but the photometric redshift solution is unconstrained).
\end{enumerate}

In Table~\ref{t-few} we list the 12 ISOCAM sources without sufficient
photometric data for a reliable SED-fitting analysis. In Col.~1 we
list the ISOCAM source identification, in Col.~2 the identification of
the optical/NIR counterpart in the LPS catalogue, in Col.~3 the
spectroscopic redshift, if available, and its reference.

\begin{table}
\centering
\caption[]{Additional ISOCAM sources}
\label{t-few}
\begin{tabular}{lll}
\hline
ISOCAM & Optical         & Redshift \\
source id. & counterpart     & (reference) \\
\hline
ISO\_A2218\_07  & L599 & -- \\   
ISO\_A2218\_15a & L515 & -- \\   
ISO\_A2218\_37  & L292 & -- \\   
ISO\_A2218\_41  & L283 & -- \\   
ISO\_A2218\_46  & L242 & 0.654 (E98) \\   
ISO\_A2218\_49  & L225 & -- \\   
ISO\_A2218\_51  & L220 & -- \\
ISO\_A2218\_64  & L111 & -- \\   
ISO\_A2218\_65  & L96  & 0.64 (Paper~I) \\   
ISO\_A2218\_69  & L60  & -- \\   
ISO\_A2218\_74  & L33  & -- \\
ISO\_A2218\_75  & L26  & 0.1688 (LPS) \\   
\hline
\end{tabular}
\end{table}

In summary (see also Table~\ref{t-sum}), we obtain acceptable fits for
41 out of the 48 galaxy SEDs, five of which have no spectroscopic-$z$
estimate. Of these five, four have photometric-$z$'s consistent with
the mean cluster $z$, so we consider them as cluster members.  In
total, acceptable fits are obtained for 27 cluster members and 14
field galaxies. For two sources the best fit is unconstrained because
of a degeneracy problem (different models at very different $z$'s
provide fits of comparable quality). For another five sources, no
acceptable fit was found (they are discussed in
Appendix~\ref{a-sources}).

\begin{table}
\centering
\caption[]{Summary of the SED fitting analysis}
\label{t-sum}
\begin{tabular}{lr}
\hline
Sample & number \\
\hline
Non-stellar sources & 60 \\
Sources with well-sampled SED & 48 \\
Sources without acceptable SED model fit & 5 \\
Sources with unconstrained SED model fit & 2 \\
Sources with acceptable SED model fit & 41 \\
Cluster sources with acceptable SED model fit & 27 \\
Field sources with acceptable SED model fit & 14 \\
\hline
\end{tabular}
\end{table}

\section{The IR Luminosities and Star Formation Rates}\label{s-sfr}
Computing the IR luminosities, $L_{IR}$'s, of our MIR sources requires
knowledge of both their redshifts and SEDs. We can therefore determine
the $L_{IR}$'s for 41 ISOCAM sources in the A2218 field.

We adopt two methods to determine $L_{IR}$: (a) We integrate the
best fit model SED over the range 8--1000 $\mu$m; (b) we use the
empirical relations of Elbaz et al.  (\cite{elba02}, eqs. 13 and 14 in
that paper), between $L_{IR}$ and the K-corrected monochromatic MIR
luminosities at 6.7 and 14.3 $\mu$m. The best-fit model SED is used to
estimate the K-correction. In both cases, the luminosities of
background sources are corrected for the lensing amplification factors
(taken from Tables 5 and 8 of Paper~I).

\begin{figure}
\centering
\resizebox{\hsize}{!}{\includegraphics{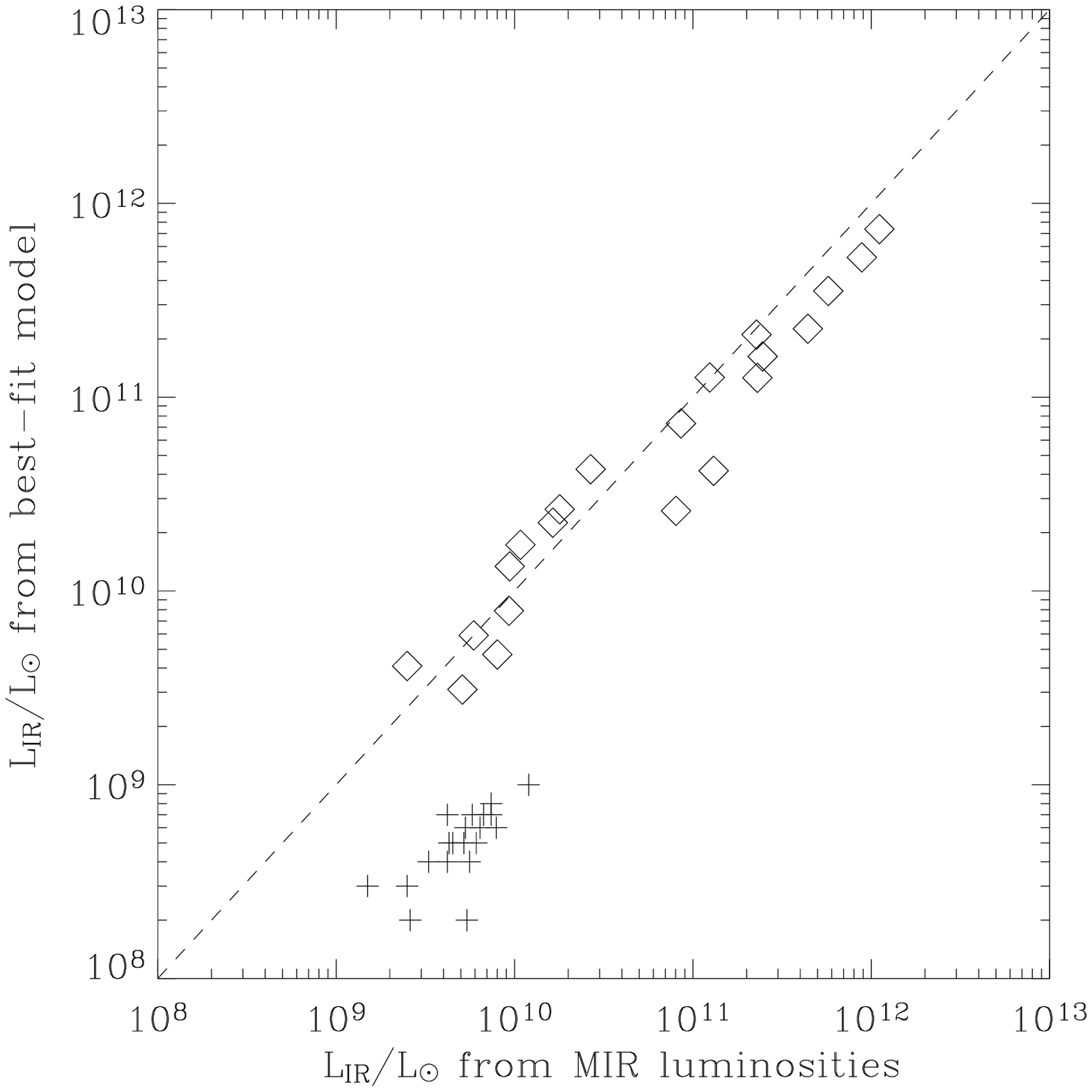}}
\caption{The ISOCAM sources IR luminosities obtained from the
best-fit model SEDs plotted vs. the IR luminosities obtained from
the K-corrected LW3 luminosities, or, when these are not available,
from the K-corrected LW2 luminosities, through
the relations of Elbaz et al. (\cite{elba02}).
The dashed line is the identity line. Crosses indicate those
sources whose SED is best fit by an E-type model.}
\label{f-lircmp}
\end{figure}

The $L_{IR}$ estimates obtained by the two methods are very similar,
except for those sources with SEDs that are best fit by E-type models,
for which method (b) returns systematically higher luminosities than
method (a) -- see Fig.~\ref{f-lircmp}. This is not surprising, since
the relations of Elbaz et al.  (\cite{elba02}) are consistent with
those of Chary \& Elbaz (\cite{char01}), which were derived on a
sample of galaxies with significant star formation activity.  The SEDs
of the galaxies in Chary \& Elbaz's (\cite{char01}) sample are clearly
different from those of E-type galaxies, where the MIR emission is
dominated by the Rayleigh-Jeans tail of the cold stellar component
(see, e.g., Boselli et al. \cite{bose98}), not by thermal emission
from dust. Because of this problem, in the following we adopt the
$L_{IR}$-estimates obtained with method (a). We also calculate the
NIR-luminosities, $L_{NIR}$, using the same method, i.e. by
integrating the best fit model SED over the range 0.9--3.0 $\mu$m.

We then compute the SFRs from Kennicutt's (\cite{kenn98}) relation
between a galaxy SFR and its $L_{IR}$,
\begin{equation}
\mbox{SFR} (M_{\odot} \, \mbox{yr}^{-1})= 1.71 \times 10^{-10} \, L_{IR}/L_{\odot}.
\label{e-sfr}
\end{equation}

\begin{table}
\centering
\caption[]{Luminosities and SFRs}
\label{t-lums}
\begin{tabular}{lrrr}
\hline
\hline
ISOCAM & $L_{NIR}$ & $L_{IR}$ & SFR \\
source id. & $10^9 L_{\odot}$  & $10^9 L_{\odot}$ & $M_{\odot} \, {\mbox yr}^{-1}$ \\
\hline
\multicolumn{4}{c}{Cluster sources} \\
\hline
{\em ISO\_A2218\_05}   &   7.2 &    17.3 &    2.9 \\
ISO\_A2218\_06   &  58.2 &     0.7 &    0.1 \\
ISO\_A2218\_14   &  42.8 &     0.8 &    0.1 \\
{\em ISO\_A2218\_17}   &   1.4 &     4.1 &    0.7 \\
ISO\_A2218\_18   &  44.1 &     0.6 &    0.1 \\
ISO\_A2218\_21   &  18.8 &     0.2 &    0.0 \\
ISO\_A2218\_22   &  27.1 &     0.3 &    0.1 \\
ISO\_A2218\_23   &  32.3 &     0.7 &    0.1 \\
ISO\_A2218\_24   &  31.2 &     0.6 &    0.1 \\
ISO\_A2218\_25   &   9.7 &     0.2 &    0.0 \\
{\em ISO\_A2218\_26}   &   2.5 &     5.9 &    1.0 \\
ISO\_A2218\_30   &  55.0 &    26.4 &    4.5 \\
ISO\_A2218\_32   &  15.6 &     0.3 &    0.0 \\
ISO\_A2218\_33   &  40.2 &     0.5 &    0.1 \\
ISO\_A2218\_34   &   6.4 &     3.1 &    0.5 \\
ISO\_A2218\_36b  &  27.0 &     0.5 &    0.1 \\
ISO\_A2218\_39   &  24.8 &     0.4 &    0.1 \\
ISO\_A2218\_45   &  83.4 &     1.0 &    0.2 \\
ISO\_A2218\_47   &  53.6 &     0.7 &    0.1 \\
ISO\_A2218\_54   &  37.3 &     0.5 &    0.1 \\
ISO\_A2218\_57   &  50.7 &     0.6 &    0.1 \\
ISO\_A2218\_58   &  42.7 &     0.5 &    0.1 \\
{\em ISO\_A2218\_61a}  &  16.5 &     7.9 &    1.3 \\
ISO\_A2218\_61b  &  36.2 &     0.7 &    0.1 \\
ISO\_A2218\_62   &  33.6 &     0.4 &    0.1 \\
ISO\_A2218\_66   &  23.9 &     0.4 &    0.1 \\
ISO\_A2218\_70   &   9.8 &     4.7 &    0.8 \\
\hline
\multicolumn{4}{c}{Field sources} \\
\hline
ISO\_A2218\_08   &  23.6 &   353.3 &   60.1 \\
ISO\_A2218\_09   &  25.5 &   210.3 &   35.7 \\
{\em ISO\_A2218\_13}   &  41.6 &   526.6 &   89.5 \\
ISO\_A2218\_20   &   6.1 &    42.5 &    7.2 \\
ISO\_A2218\_27   &   5.6 &    13.4 &    2.3 \\
ISO\_A2218\_29   &   7.9 &    73.2 &   12.4 \\
ISO\_A2218\_38   & 103.5 &   736.5 &  125.2 \\
ISO\_A2218\_40   &   2.4 &    25.9 &    4.4 \\
ISO\_A2218\_42   &  11.8 &   125.9 &   21.4 \\
ISO\_A2218\_44   &   4.5 &    41.7 &    7.1 \\
ISO\_A2218\_52   &  38.9 &   225.6 &   38.3 \\
ISO\_A2218\_59   &  22.8 &   162.3 &   27.6 \\
ISO\_A2218\_60   &   9.4 &    22.5 &    3.8 \\
ISO\_A2218\_68   &  53.0 &   126.4 &   21.5 \\
\hline
\hline
\end{tabular}
\end{table}

In Table~\ref{t-lums} we list in Col.~1 the source identification
(sources without spectroscopic redshift are listed in italic), in
Cols.~2 and 3, $L_{NIR}$ and $L_{IR}$, respectively, in units of $10^9
L_{\odot}$, and finally in Col.~4 the SFR, in units of
$M_{\odot} \, {\mbox yr}^{-1}$. Only the 27 cluster sources and the 14 field
sources with acceptable model fits to their SEDs are listed.

\section{Cluster sources}\label{s-cl}
\subsection{Internal properties}\label{ss-clint}
\begin{figure*}
\centering
\resizebox{\hsize}{!}{\includegraphics{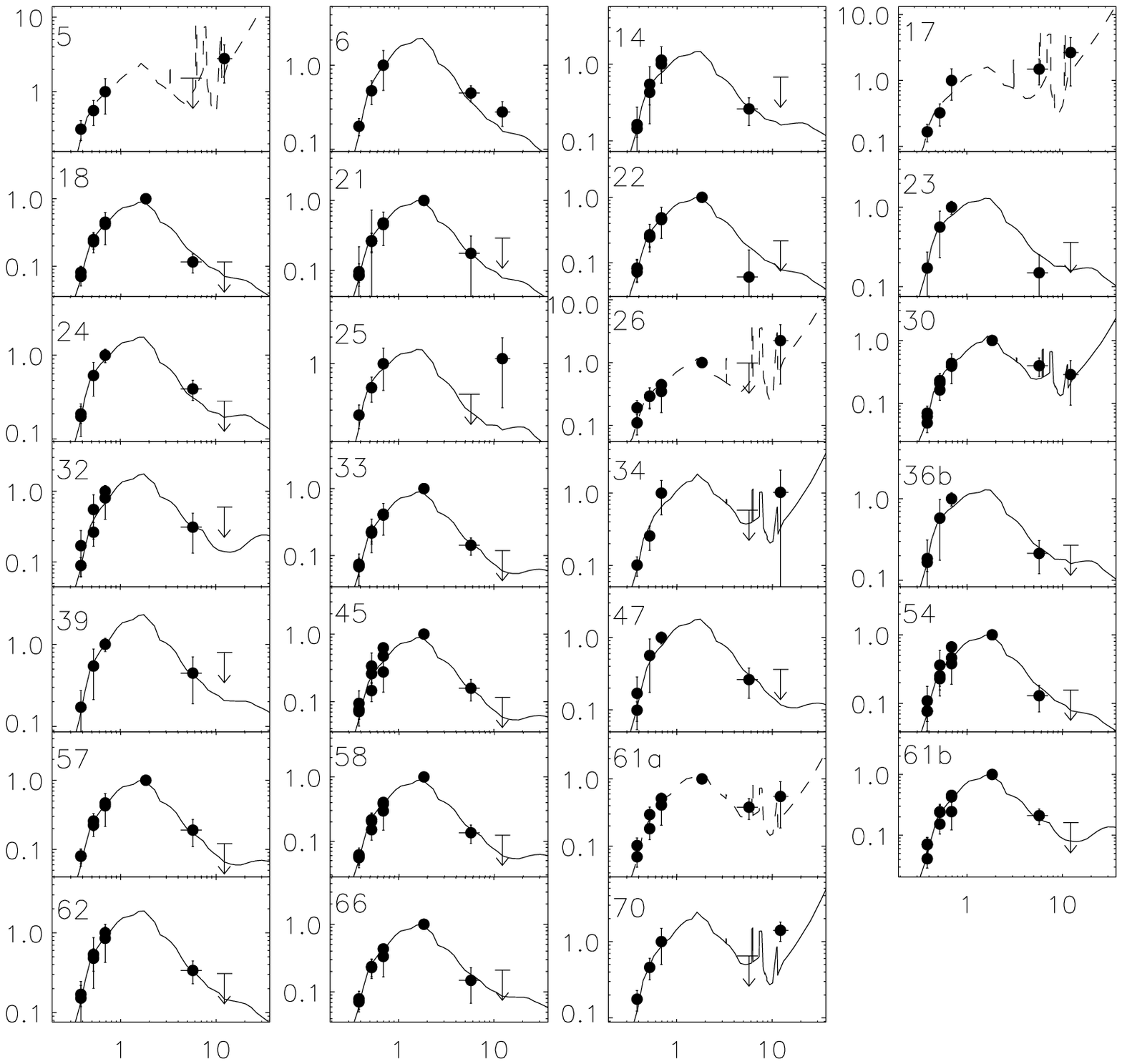}}
\caption{The observed SEDs of the 27 cluster sources for which we
obtain an acceptable fit (dots with 2-$\sigma$ error bars), with the
best-fit GRASIL models overlayed (solid lines for sources with
spectroscopic $z$, dashed line for sources with $z$ determined from
the model fitting procedure). The rest-frame wavelength, in $\mu$m, is
plotted along the x-axis, and the flux density (in normalized units)
is plotted along the y-axis. Upper limits are indicated by arrows.
The approximate widths of the LW2 and LW3 bands are indicated.}
\label{f-sed_cl}
\end{figure*}

In Fig.~\ref{f-sed_cl} we show the SEDs of the 27 cluster sources for
which an acceptable fit was found (see Table~\ref{t-sum}), with the
best-fitting models overlayed. Most cluster sources are characterized
by a SED that decreases from the NIR to the MIR, typical of passively
evolving galaxies. Models with $\mbox{Nsfr(0.1)} \simeq
\mbox{Nsfr(1.1)} \simeq 0$ provide the best-fit to the SEDs of 20
cluster sources (see Fig.~\ref{f-fstar}, middle panel). Models with
slightly higher values of Nsfr(0.1) and Nsfr(1.1) (but still below
unity) provide the best-fit to the SEDs of the remaining seven cluster
sources (three of which however lack spectroscopic redshift
determinations, so their cluster membership is not certain).  Models
with $\mbox{Nsfr(0.1)} \simeq \mbox{Nsfr(1.1)} \simeq 0$ are still
acceptable for four of these seven galaxies (see Table~\ref{t-sum}),
and only three of them have Nsfr(0.1) significantly greater than zero,
implying significant, albeit small, ongoing star formation.

\begin{figure}
\centering
\resizebox{\hsize}{!}{\includegraphics{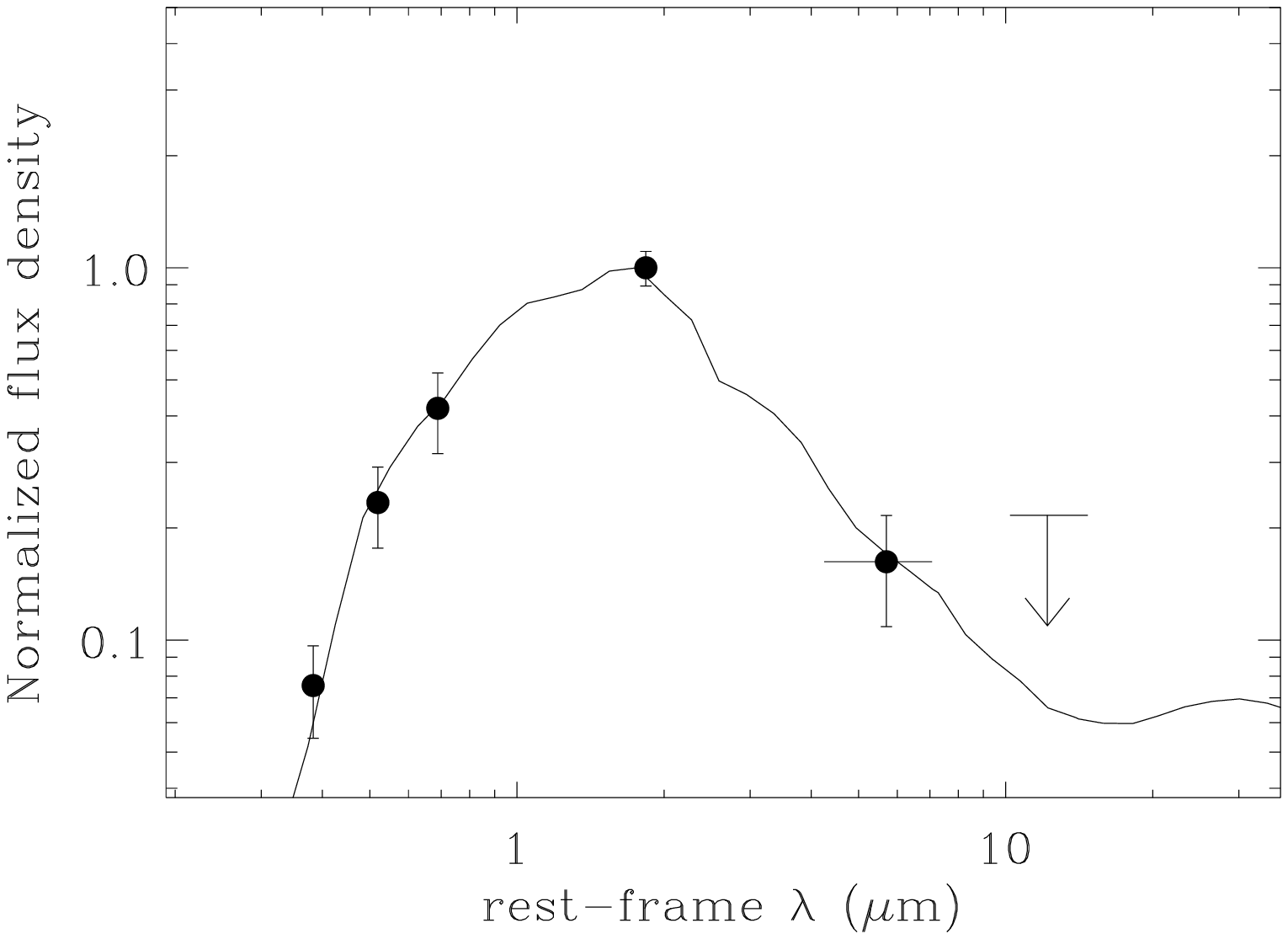}}
\caption{The average SED of the 27 ISOCAM cluster sources. Error-bars
denote the rms among the flux values. The arrow indicates the upper
limit on the LW3 flux value. The approximate widths of the LW2 and
LW3 bands are indicated. The 27 SEDs were normalised
to the $H$-band flux density, as estimated from the individual best
fit models, before averaging them.  The best fit Em10 model SED is
also shown (solid line).}
\label{f-avesed}
\end{figure}

In order to make a more efficient use of our data, we combine the 27
SEDs of the ISOCAM cluster sources. We assume the mean cluster
redshift for all cluster sources, and we scale the observed SED of
each source by the rest-frame $H$-band flux, as estimated from the
best fit model. Not all sources are detected in the LW3 ISOCAM band.
We estimate the average LW3 flux using both the measured LW3 flux
values and the LW3 flux upper limits. The average LW3 flux we derive
is therefore an overestimate, and we treat it as an upper limit. The
resulting average SED is shown in Fig.~\ref{f-avesed}, with the best
fit model, an Em10, superposed. This model (and all acceptable
models, at $\geq 5$\% c.l.) has
$\mbox{Nsfr(0.1)}=\mbox{Nsfr(1.1)}=0$. This analysis therefore
confirms that, on average, ISOCAM cluster sources have negligible
ongoing, or even recent, star formation. This result indicates they
must have formed the bulk of their stars quite some time before the
observing epoch.

\begin{figure}
\centering
\resizebox{\hsize}{!}{\includegraphics{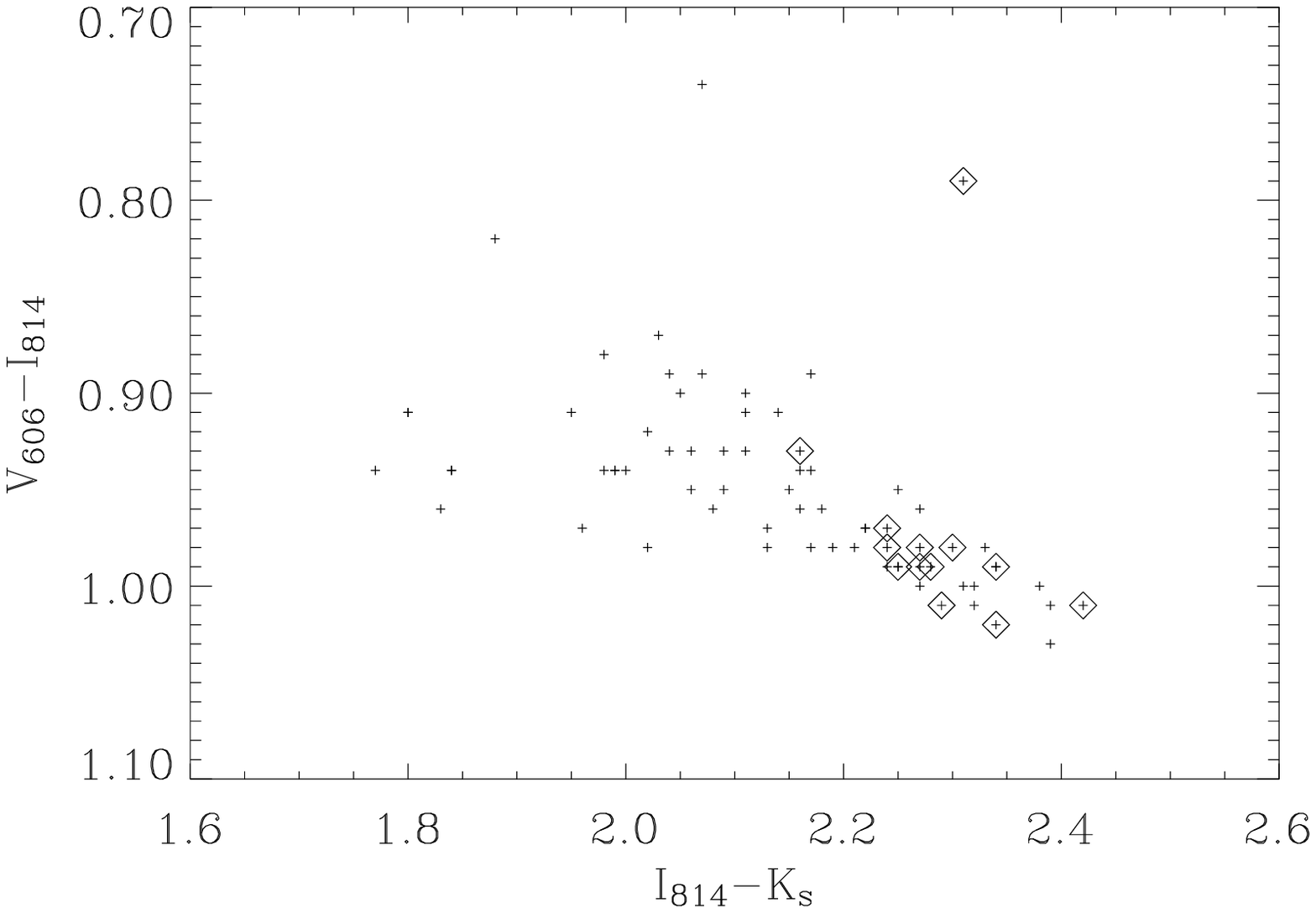}}
\caption{The $V_{606}-I_{814}$ vs. $I_{814}-K_s$ colour-colour 
diagram for cluster sources with data from S01. ISOCAM galaxies
are plotted as diamonds.}
\label{f-colcol}
\end{figure}

Our conclusion is in agreement with what can be inferred from optical
and NIR data. In fact, morphologically, most ISOCAM cluster sources
are classified as ellipticals or S0's, which in general are characterized
by old stellar populations. A similar conclusion can be reached from
the analysis of their optical/NIR colours. In Fig.~\ref{f-colcol} we
plot the $V_{606}-I_{814}$ vs. $I_{814}-K_s$ colour-colour diagram from
the cluster galaxies in S01's sample. ISOCAM sources lie in the
sequence defined by the redder (and more luminous) galaxies, except
source 26, a galaxy with an Sdm optical morphology (whose cluster
membership, as mentioned before, is not certain). On the basis of this
colour-colour diagram, S01 determine ages from 5 to 10 Gyr for the
redder and more luminous A2218 cluster members (see Figure~3 in S01).

The consistent age estimates derived from the optical and the IR data
suggest that there is little dust, on average, in the A2218 cluster
galaxies, and that the MIR emission of ISOCAM cluster sources comes
from the Rayleigh-Jeans tail of the photospheric emission from cold
stars (see, e.g., Boselli et al. \cite{bose98})

\begin{figure*}
\centering
\resizebox{\hsize}{!}{\includegraphics{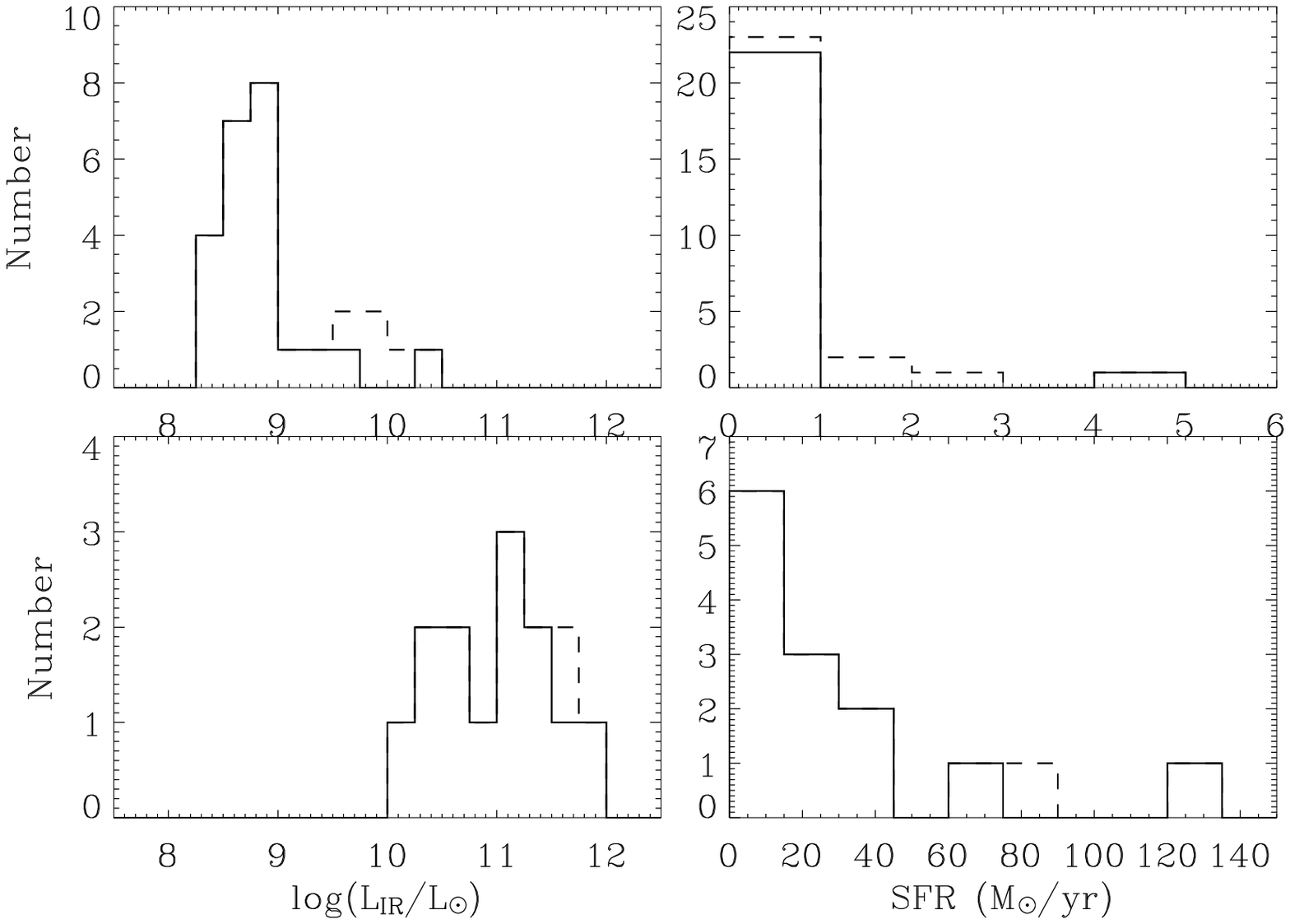}}
\caption{The distributions of $L_{IR}$ and SFR for ISOCAM cluster and
field sources. Solid (dashed) histograms show the distributions for
sources with (respectively, without) known redshift.  Top left panel:
the distribution of $L_{IR}$ for cluster sources.  Bottom left panel:
the distribution of $L_{IR}$ for field sources.  Top right panel: the
distribution of SFR for cluster sources.  Bottom right panel: the
distribution of SFR for field sources.}
\label{f-lirsfr}
\end{figure*}

The $L_{IR}$- and SFR-distribution of cluster sources are shown in the
two top panels of Fig.~\ref{f-lirsfr}. The average $L_{IR}$ is $3
\times 10^9 L_{\odot}$, but the median is only $6 \times 10^8
L_{\odot}$, and in fact most cluster sources have $L_{IR} \leq 10^9
L_{\odot}$, corresponding to SFR $<1 M_{\odot} \, {\mbox
yr}^{-1}$. There are a few sources with moderate $L_{IR}$'s and SFRs,
but none qualify as a Luminous IR Galaxy (LIRG, $L_{IR} \geq 10^{11}
L_{\odot}$, see Genzel \& Cesarsky \cite{genz00}). Source 30 has the
highest SFR, but even in this case it is only $\sim 4 M_{\odot}
\, {\mbox yr}^{-1}$.

\begin{figure}
\centering
\resizebox{\hsize}{!}{\includegraphics{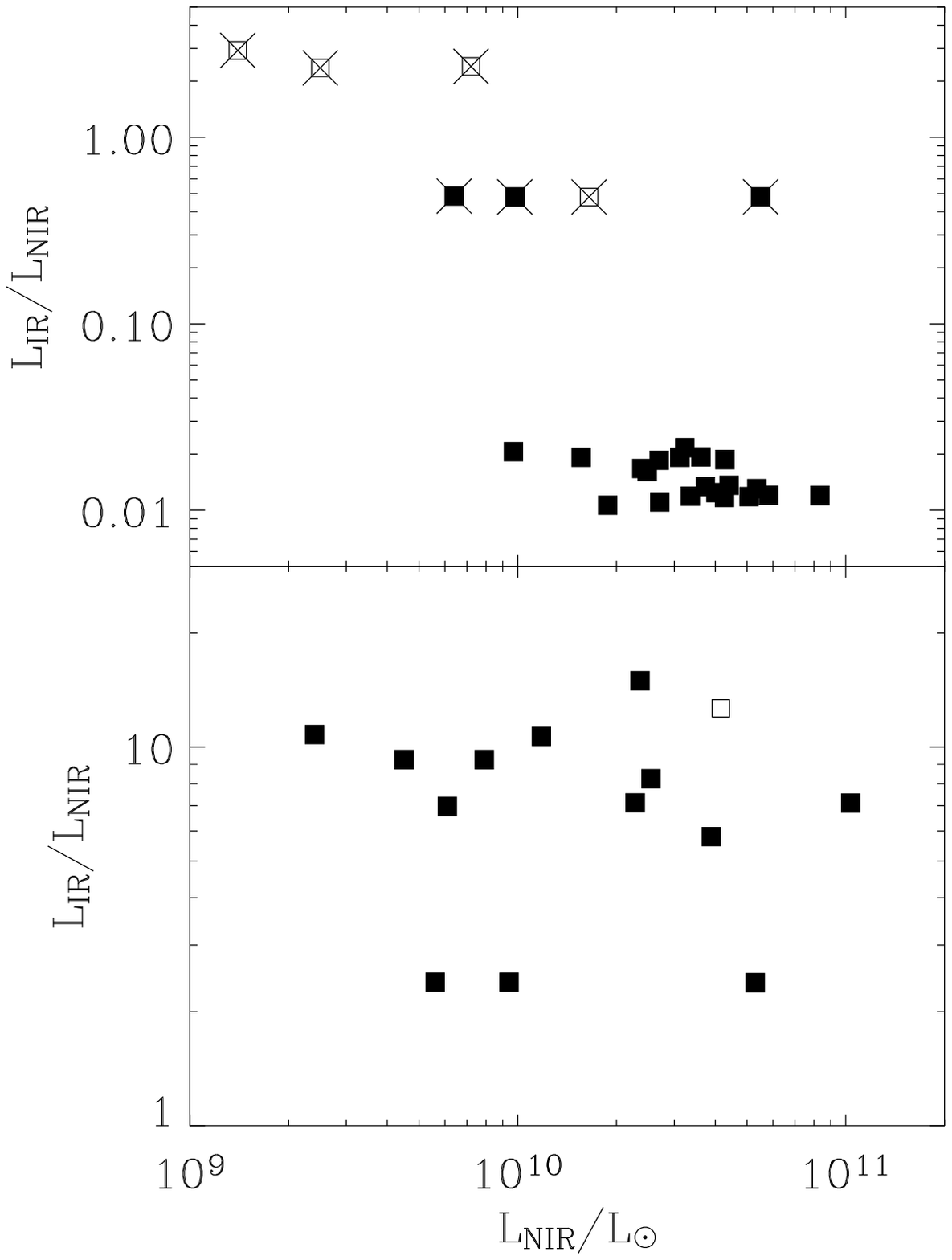}}
\caption{The ratio of the IR luminosity to the NIR luminosity,
$L_{IR}/L_{NIR}$ as a function of $L_{NIR}$, for ISOCAM sources.  Open
symbols indicate sources without spectroscopic redshift estimates. Top
panel: cluster members. Sources with Nsfr(0.1) $>0$ are
crossed.  Bottom panel: field galaxies.}
\label{f-lirlnir}
\end{figure}

If the galaxy is small, even a relatively small SFR can be considered
a sign of strong ongoing star formation activity. In order to
investigate this, we consider the $L_{IR}/L_{NIR}$ ratio
vs. $L_{NIR}$. The SFR scales linearly with $L_{IR}$
(eq.~\ref{e-sfr}), and $L_{NIR}$ is a fair indicator of the total
baryonic mass of a galaxy (e.g. Gavazzi et al.  \cite{gava96}), so
$L_{IR}/L_{NIR}$ is proportional to the SFR normalised by the baryonic
mass of a galaxy.  We plot $L_{IR}/L_{NIR}$ vs. $L_{NIR}$ for cluster
sources in Fig.~\ref{f-lirlnir} (top panel).  For most cluster sources
$L_{IR}/L_{NIR} <0.03$, and for all spectroscopically-confirmed
cluster members $L_{IR}/L_{NIR} <0.5$. Only three
non-spectroscopically confirmed cluster members (sources 5, 17, and
26) have $L_{IR} > L_{NIR}$, and for one of them (source 5) the
assigned optical counterpart is not certain (see \S~\ref{s-data}). Thus,
in general, even the normalised SFRs of ISOCAM cluster sources are
small, although the SFR per unit baryonic mass seems to be higher for
galaxies of lower NIR luminosity.

We conclude that most ISOCAM cluster sources are passively evolving
galaxies. Their MIR emission is of stellar photospheric origin. There
are nonetheless ISOCAM cluster sources with non-zero, albeit small,
SFRs, and the most active of these sources are those with the smallest
baryonic masses.

\subsection{Spatial and velocity distributions}\label{ss-cldist}
\begin{figure}
\centering
\resizebox{\hsize}{!}{\includegraphics{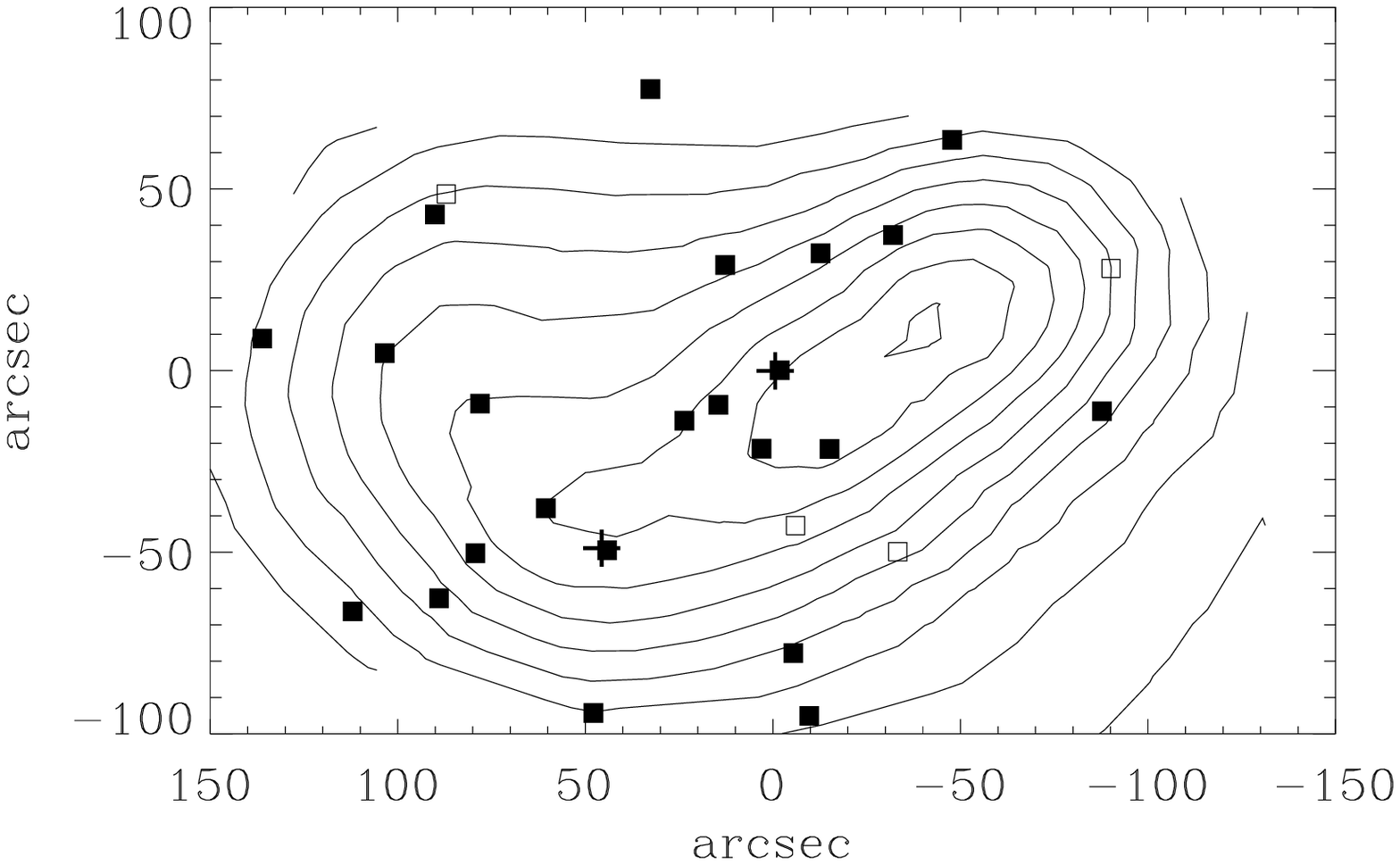}}
\caption{The projected space distribution of ISOCAM cluster sources
(filled squares: spectroscopically confirmed members; open squares:
sources whose cluster membership is established from the best fit
model SED), and of optically-selected cluster sources (isodensity
contours). The two crosses indicate the position of the cD (near the
centre) and of LPS244, the second brightest cluster galaxy.}
\label{f-k2d}
\end{figure}

In order to obtain useful information from the spatial and velocity
distributions of ISOCAM cluster members, we compare them with those of
optically selected cluster sources. In Fig.~\ref{f-k2d} we show a
smoothed projected number-density map, (obtained with the method of
the adaptive kernel, see, e.g., Biviano et al.  \cite{bivi96}), of the
266 likely cluster members from the LPS catalogue. The membership has
been established by selecting galaxies within $\pm 0.25$ magnitude of
the colour-magnitude sequence in the $B-r$ vs. $r$ diagram (see, e.g.,
Biviano et al. \cite{bivi96}). Superposed on the density map are the
ISOCAM galaxies belonging to the cluster, and with photometric data
from LPS.  The ISOCAM sources seem to share the same spatial projected
distribution of optically-selected sources, except perhaps that they
seem to avoid the main density peak. However, the relative fractions
of optically-selected and MIR-selected cluster members contained
within the highest-density cluster region (i.e. within the second
highest isocontour of Fig.~\ref{f-k2d}) are not significantly
different (23/266 and 2/27, respectively). And in fact, the whole
spatial distributions of the two samples are not significantly
different (22~\% c.l.) according to the bi-dimensional
Kolmogorov-Smirnov test (see, e.g., Fasano \& Franceschini
\cite{fasa87}).

\begin{figure}
\centering
\resizebox{\hsize}{!}{\includegraphics{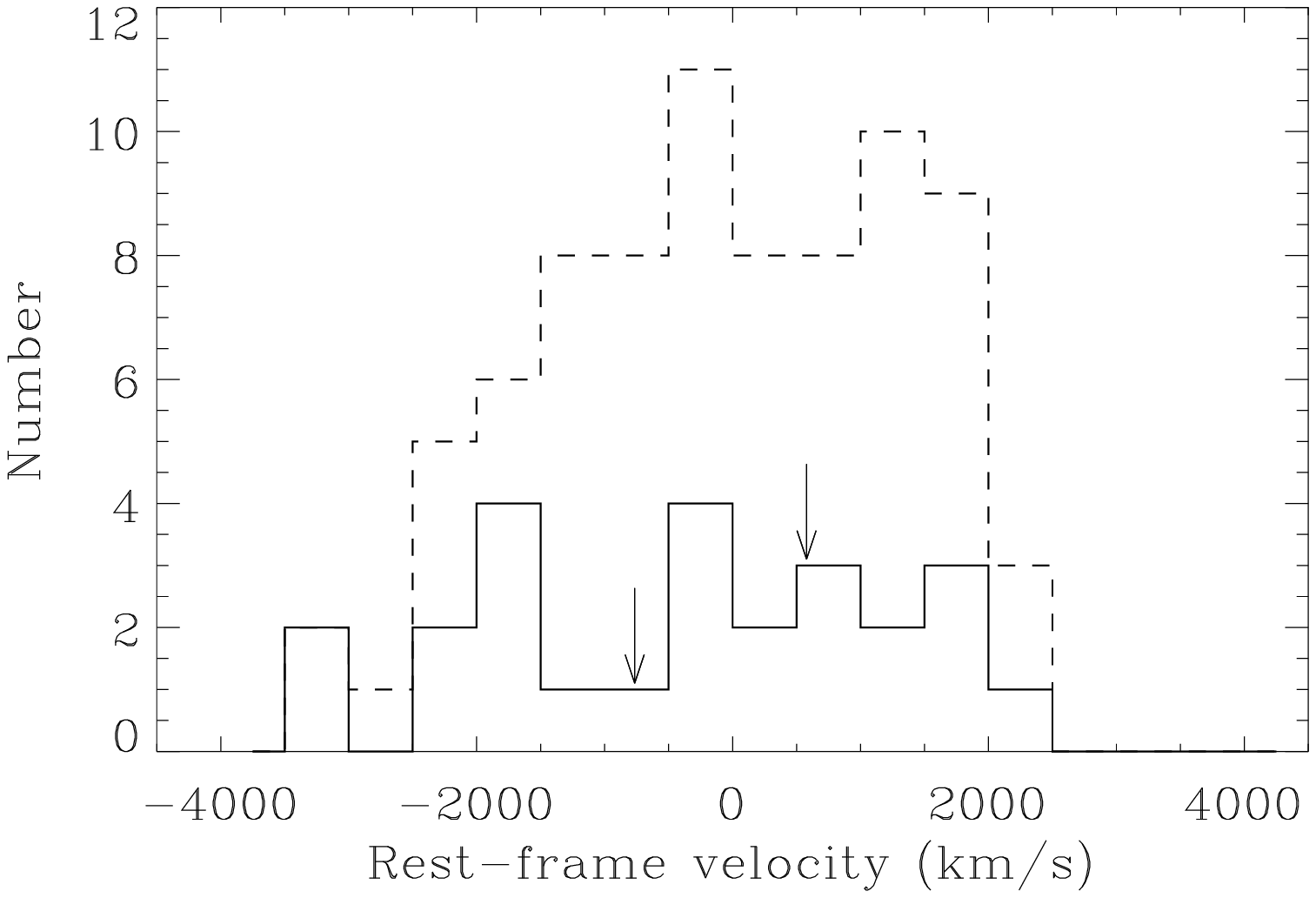}}
\caption{The rest-frame velocity distribution of optically-selected
cluster members (dashed line) and of ISOCAM cluster sources (solid
line). Arrows indicate the location in the velocity space of the cD
galaxy (at negative velocity) and the 2nd brightest member, LPS244.}
\label{f-vhisto}
\end{figure}

We now check the relative distributions of optically-selected and
MIR-selected cluster members in velocity space. We select 79
cluster members from the full sample of 101 galaxies with measured
redshifts in the A2218 field, using the method of den Hartog \&
Katgert (\cite{denh96}). The velocity distribution of the A2218
cluster members is shown in Fig.~\ref{f-vhisto} (note that velocities
are rescaled to the cluster rest-frame using the average cluster
velocity). The dashed and solid histograms refer to the 79 cluster
members, and, respectively, to the subsample of 25 ISO cluster members
with available redshift (see Tables~\ref{t-sed} and \ref{t-few}). We
run the Kolmogorov-Smirnov test to test the null hypothesis that the
two samples have the same parent velocity distribution; the null hypothesis
cannot be rejected (17~\% c.l.).

The average velocity and velocity dispersion of the 79
optically-selected cluster members (corrected for cosmological effects
and velocity measurement errors, see Danese et al.  \cite{dane80}),
are $52376 \pm 188$ km~s$^{-1}$ and $1412_{-108}^{+117}$ km~s$^{-1}$,
respectively (here and hereafter, the biweigth estimators for the
location and scale of a distribution are used, see Beers et
al. \cite{beer90}).  These values are in agreement with those given by
Cannon et al. (\cite{cann99}) and Girardi \& Mezzetti (\cite{gira01}),
and are not significantly different from the average velocity and
velocity dispersion of the 25 ISOCAM cluster members ($52091 \pm 388$
km~s$^{-1}$ and $1606_{-215}^{+247}$ km~s$^{-1}$, respectively).

We conclude that the projected phase-space distribution of
ISOCAM-selected cluster members is not significantly different
from that of the global cluster population.

It is nonetheless interesting to analyse the distribution of cluster
galaxies in some more detail, to search for the existence of
substructures. We find that the velocity distribution of the 79
central cluster members is marginally different from a Gaussian (98~\%
c.l.), according to the Anderson-Darling test (see, e.g., D'Agostino
\cite{dago86}), even if a battery of other tests (skewness, kurtosis,
tail-index, see Bird \& Beers \cite{bird93}) fail to find a
significant difference. We also note that the velocity of the cD
galaxy is significantly offset from the cluster mean (as indicated by
the test of Gebhardt \& Beers \cite{gebh91}; see also
Fig.~\ref{f-vhisto}). The non-Gaussian velocity distribution of the
cluster members, and the fact that the cluster cD is not at rest at
the bottom of the cluster potential, suggest that the cluster is not
fully dynamically relaxed (see, e.g., Girardi \& Biviano
\cite{gira03}).

We then look for the presence of substructures in the 3d-space of
positions and velocities. We do not detect significant substructure
with the classical test of Dressler \& Shectman
(\cite{dres88}). However, a detailed modelling of the
gravitational-lensing properties of the cluster has suggested the
presence of two main mass concentrations, one centered on the cD and
the other on the second brightest member, LPS244 (Kneib et
al. \cite{knei95}).  The model for the mass distribution of Kneib et
al. (\cite{knei95}) is remarkably similar to the projected galaxy
distribution shown in Fig.~\ref{f-k2d} (compare with Fig.5 in Kneib et
al. \cite{knei95}).

Since there is evidence for a bimodal cluster structure, one could ask
whether the ISOCAM cluster members belong to one structure rather
than to the cluster as a whole.  However, neither from the projected
spatial distribution (see Fig.~\ref{f-k2d}), nor from their velocity
distribution (see Fig.~\ref{f-vhisto}), is there any evidence for
significant subclustering of the ISOCAM cluster population.

\section{Field sources}\label{s-fd}
\begin{figure*}
\centering
\resizebox{\hsize}{!}{\includegraphics{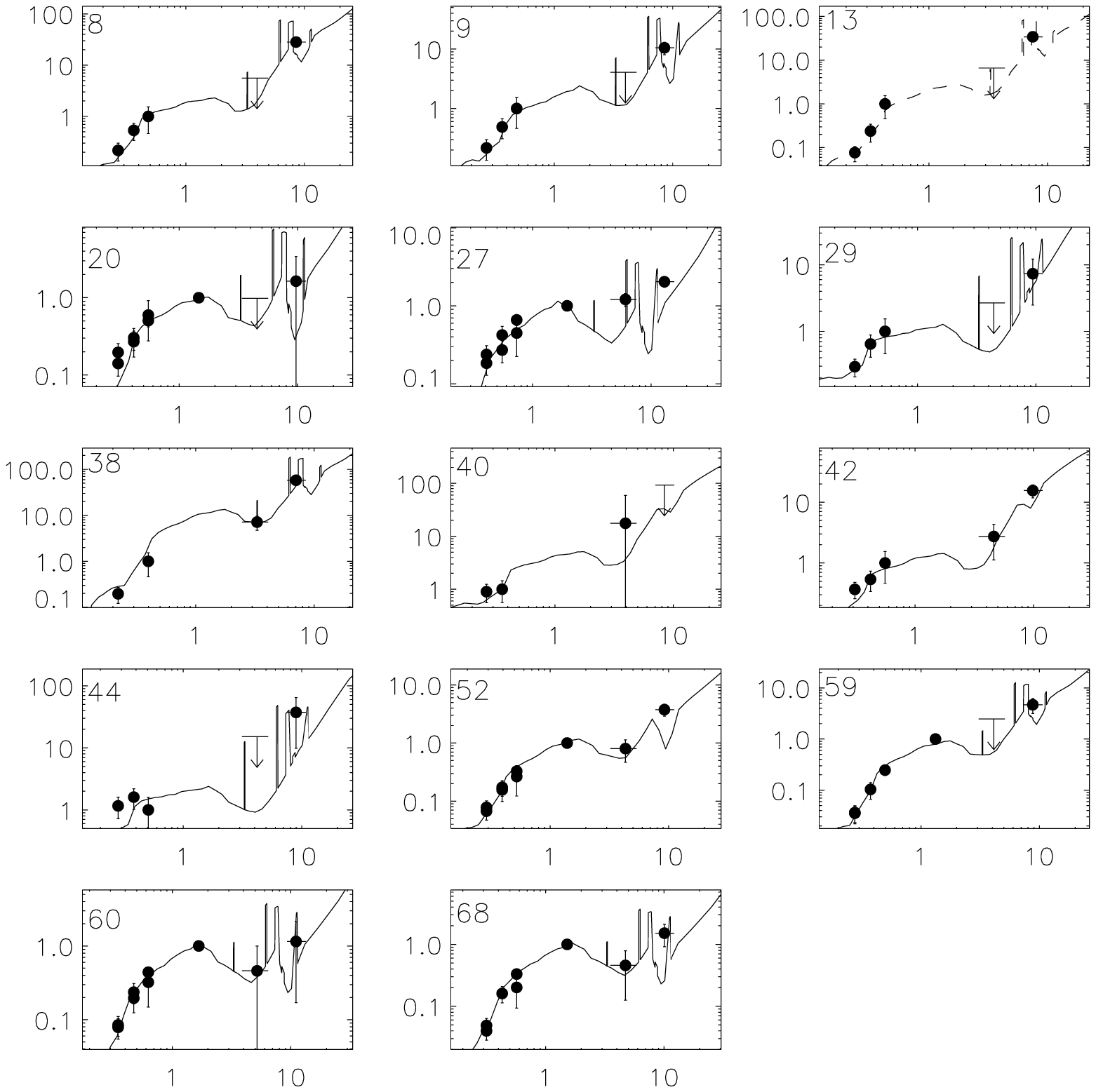}}
\caption{The observed SEDs of the 14 field sources for which we obtain
an acceptable fit (dots with 2-$\sigma$ error bars), with the best-fit
GRASIL models overlayed (solid lines for sources with spectroscopic
$z$, dashed line for the source with $z$ determined from the model
fitting procedure). The rest-frame wavelength, in $\mu$m, is plotted
along the x-axis, and the flux density (in arbitrary units) is
plotted along the y-axis. Upper limits are indicated by arrows.
The approximate widths of the LW2 and LW3 bands are indicated.}
\label{f-sed_fd}
\end{figure*}

In Fig.~\ref{f-sed_fd} we show the SEDs of the 14 field sources for
which an acceptable fit was found (see Table~\ref{t-sum}), with the
best-fitting models overlayed. The SEDs of field sources are quite
different from those of cluster sources, and are charachterized by an
increasing flux density with wavelength. Such a SED is characteristic
of actively star-forming galaxies, where part, or even most, of the
stellar radiation is reprocessed by dust and re-emitted in the IR. In
fact, models that provide the best-fits to the field-source SEDs have
median values of $\mbox{Nsfr(0.1)}=1.1$ and
$\mbox{Nsfr(1.1)}=1.4$, indicating substantial ongoing and recent star
formation, at a rate higher, on average, than the mean SFR over the
galaxy SFH.  Taking into consideration all the acceptable models, we
note that Nsfr(0.1) and Nsfr(1.1) are both significantly higher than
zero in 11 (out of 14) field sources.

The ongoing star formation activity in field sources is consistent
with them being mostly (late) spirals or irregulars, as expected
from the morphology-density relation (e.g. Dressler \cite{dres80}).

\begin{figure}
\centering
\resizebox{\hsize}{!}{\includegraphics{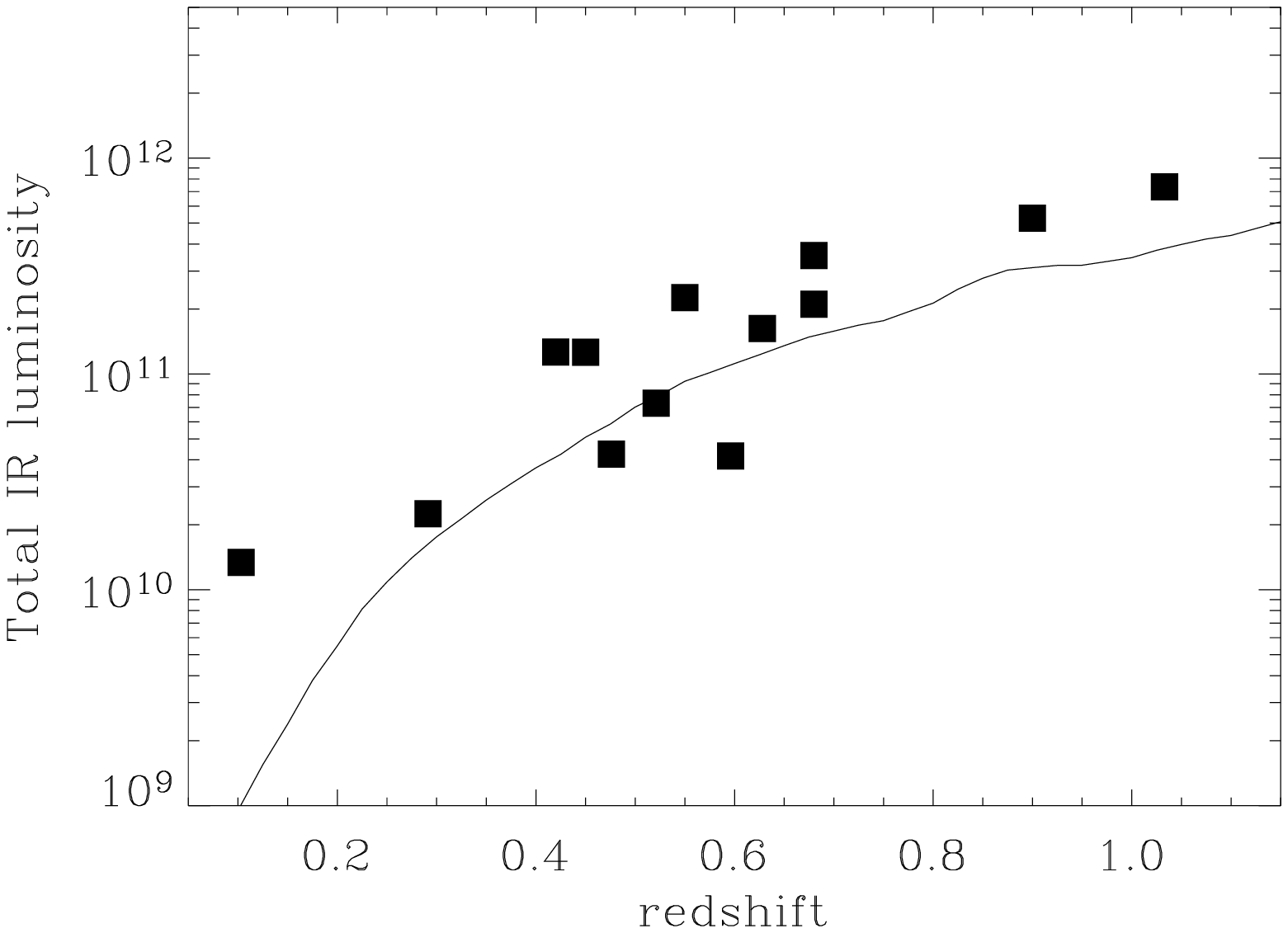}}
\caption{The total IR luminosity as a function of redshift for ISOCAM
field sources with spectroscopic redshift estimates, detected in the
LW3 band.  The solid line is the luminosity corresponding to a flux of
0.121 mJy (the 50\% completeness limit of the A2218 ISOCAM survey),
reduced by a factor 0.6 (the average lensing correction factor), and
K-corrected using the average correction for the models that best fit
the observed field galaxy SEDs.}
\label{f-ltirz}
\end{figure}

The $L_{IR}$ of field sources correlates with redshift (see
Fig.~\ref{f-ltirz}), as expected in a flux-limited survey. In
Fig.~\ref{f-ltirz} we show that the lower envelope of the observed
$L_{IR}$ vs. $z$ relation approximately corresponds to the expected
relation for a source with an observed 14.3~$\mu$m flux density 0.121
mJy (the 50\% completeness limit of the A2218 ISOCAM survey, see
Paper~I), corrected by the average amplification factor (0.6) due to
the cluster lensing, and using the average K-corrections for the
models that best fit the observed field source SEDs.

The $L_{IR}$- and SFR-distribution of field sources are shown in
Fig.~\ref{f-lirsfr} (bottom panels). The average $L_{IR}$ is $1.9
\times 10^{11} L_{\odot}$, and the median is $1.3 \times 10^{11}
L_{\odot}$. Eight of the 14 field sources classify as
LIRGs ($\sim 60$\% of all field sources). All LIRGs are sources with
known redshift. These high $L_{IR}$'s translate (via eq.~\ref{e-sfr})
into high SFRs, 2--125 $M_{\odot} \, \mbox{yr}^{-1}$, with an average
(median) of 33 (22) $M_{\odot} \, \mbox{yr}^{-1}$.  The highest-SFR
source in our sample is no.38, a lensed object at $z=1.033$ already
detected in the MIR by Barvainis et al. (\cite{barv99}).

The distribution of the field sources in the Nsfr(1.1)
vs. Nsfr(0.1) diagram (see Figure~\ref{f-fstar}, bottom panel) can help
us understand better the nature of these field sources. Most of them
are located in the central part of the diagram, characterized by
similar values of Nsfr(1.1) and Nsfr(0.1), both above unity (see also
Table~\ref{t-sed}), and only for one source is Nsfr(0.1) significantly
higher than Nsfr(1.1). This suggests that we are not observing these
galaxies during an exceptional starburst event, and that their current
SFR was maintained at a similar level for quite some time. If
anything, the median value of Nsfr(1.1) of these 14 field galaxies is
higher than their median value of Nsfr(0.1), implying a higher SFR in
the recent past than today. Consistently, almost half of the galaxies
in the field sample are located in the Nsfr(0.1)-Nsfr(1.1) part of the
diagram also occupied by PSB models (see Figure~\ref{f-fstar}, bottom
panel). However, on a galaxy per galaxy basis, the values of Nsfr(0.1)
and Nsfr(1.1) are not significantly different, and other models,
implying no major starbursts (e.g. intermediate-age spirals) fit the
field galaxy SEDs as well as the PSB models. 

With the present data we are therefore unable to say whether the
IR-selected field galaxies in our sample experienced one (or several)
starburst events in the recent past, or whether they underwent a
smoother evolution. However, the SED fitting analysis suggests that
their SFRs have not changed significantly in the last $\sim 1$ Gyr.
This conclusion implies that, on average, $\sim 3 \times 10^{10} \,
M_{\odot}$ of stars were produced in the last Gyr before the observing
epoch. This represents a substantial fraction of the total baryonic
mass of these galaxies (as estimated from the best-fit GRASIL models)
$\sim 30$\%, on average.

These ISOCAM field sources are therefore powerful star-forming
galaxies, and they shine in the IR because a substantial fraction of
their stellar luminosity is re-emitted in the IR. The amount of
extinction is $\sim 0.5$--1.5 and $\sim 1$--2 magnitudes,
respectively, in the rest-frame $B$ and $U$ bands, for the typical
best-fit models. All field sources have $L_{IR} > 2 L_{NIR}$, and a
few have $L_{IR}/L_{NIR} > 10$ (see Fig.~\ref{f-lirlnir}, bottom
panel).

\section{The Butcher-Oemler effect}\label{s-bo}
\begin{figure}
\centering
\resizebox{\hsize}{!}{\includegraphics{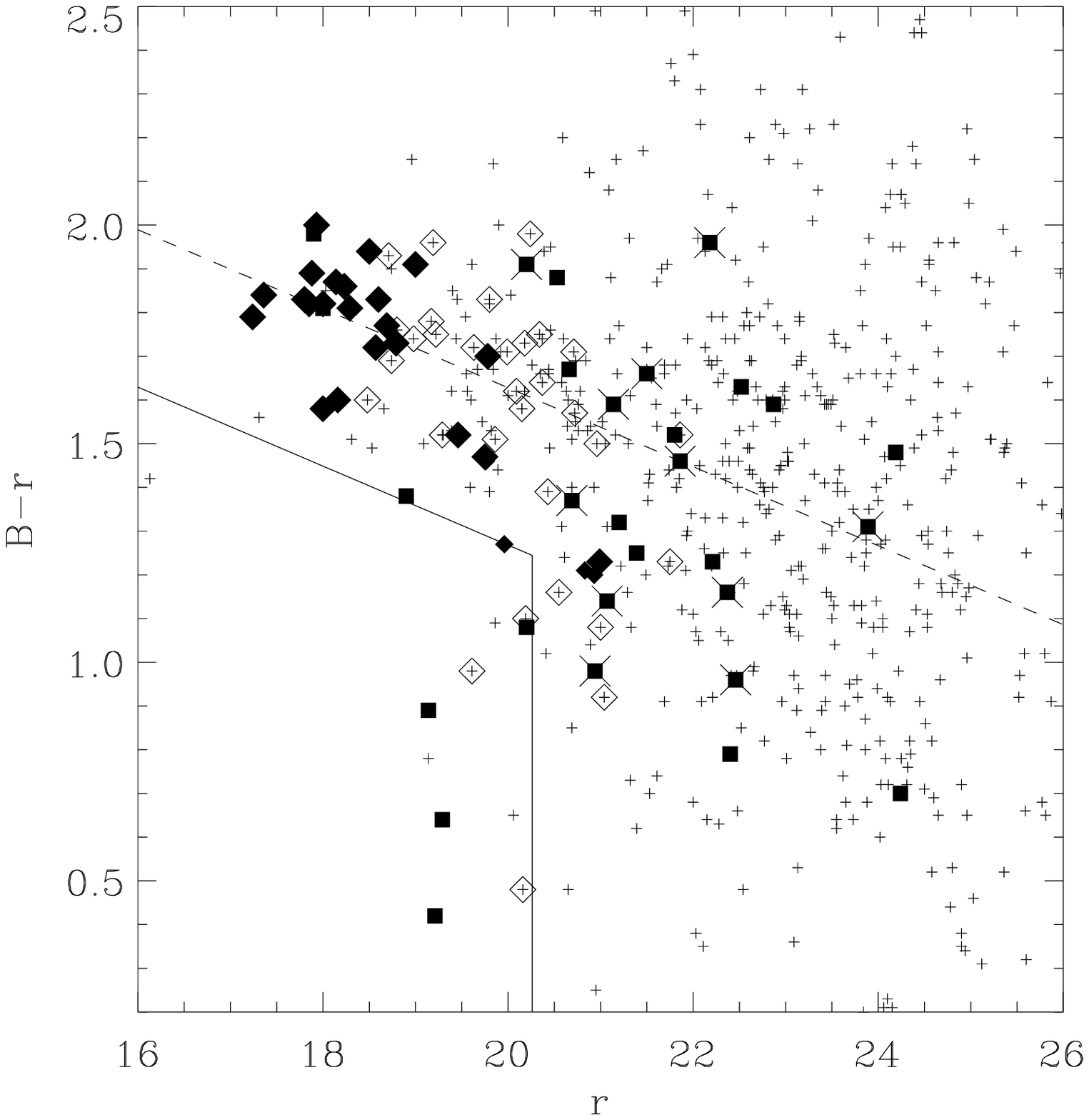}}
\caption{The $B-r$ vs. $r$ colour-magnitude diagram for galaxies in
the A2218 field (magnitude and colour data are from LPS). Filled
symbols represent ISOCAM sources.  Diamonds represent cluster sources
(big diamonds: spectroscopically confirmed cluster members; small
diamonds: cluster membership established through the SED-fitting
procedure). X's identify the ISOCAM sources with a SFR $> 4
M_{\odot}$~yr$^{-1}$.  The dashed line is the best-fit cluster
colour-magnitude relation.  Solid lines delimit the region of the
diagram where blue BO galaxies are located.}
\label{f-cmr}
\end{figure}

In Fig.~\ref{f-cmr} we show the $B-r$ vs. $r$ colour-magnitude diagram
for all the sources in the LPS catalogue. Among these, the sources
detected by ISOCAM are plotted as filled symbols, and confirmed
cluster sources are plotted as diamonds. The dashed line is a best-fit
to the main sequence of cluster members. All ISOCAM cluster sources
lie close to the main colour-magnitude sequence.

The fraction of blue BO galaxies was defined by Butcher \& Oemler
(\cite{butc84}) as the fraction of galaxies 0.2 mag bluer than the
early-type galaxies, and brighter than a K-corrected (but not
evolutionary corrected) $M_V=-20$ (they used
$H_0=50$~km~s$^{-1}$~Mpc$^{-1}$, and $q_0=0.1$). This absolute
magnitude limit corresponds to an apparent magnitude limit $r=20.27$,
after correcting for the Galactic absorption in the direction of A2218
(as given by Schlegel et al. \cite{schl98}), and for the K-correction
(we take the values from Poggianti \cite{pogg97}), and using the $V-r$
color of an early-type galaxy at $z \sim 0.175$ (Fukugita et
al. \cite{fuku95}). At the cluster redshift, a $B-V$ colour difference
of 0.2 mag corresponds roughly to the difference between the colours
of an E and a S0 galaxy; at the same redshift, the $B-r$ colours of
the same galaxy types differ by $\simeq 0.3$ mag (Fukugita et
al. \cite{fuku95}). Therefore, in order to compute the BO effect in
A2218, we consider the galaxies with $r \leq 20.27$, and, among these,
those bluer by $\geq 0.3$ mag than the colour-magnitude sequence in
the $B-r$ vs. $r$ colour-magnitude diagram.

Using the LPS data we estimate a BO galaxy fraction $f_B=0.13 \pm
0.04$, intermediate between the value given by Butcher \& Oemler
(\cite{butc84}), $f_B = 0.09 \pm 0.03$, and that given by Rakos et
al. (\cite{rako01}), $f_B=0.23$. However, if we only consider the
confirmed cluster members, the fraction of blue BO galaxies
drastically decreases to $f_B=0.07 \pm 0.04$. Similar fractions of
blue galaxies are found when selecting ISOCAM galaxies. When we
consider all ISOCAM galaxies, the blue fraction is $f_B=0.16 \pm
0.08$; only five ISOCAM galaxies contribute to the blue fraction in
the A2218 field (sources 2, 15a, 27, 61a, and 64). Of these five
galaxies, two are in the field, and three are at unknown redshift,
although one, sources 61a, is a likely cluster member, based on the
results of the SED fit analysis.

In A2218 the BO effect is the same when galaxies are selected in the
optical or in the MIR. Hence we would conclude that there is no
MIR-BO effect in A2218. However, we might have underestimated the
size of the effect because of the relatively small field covered by
ISOCAM observations. Ellingson et al. (\cite{elli01}) have in fact
shown that in order to detect the BO effect it is necessary to sample
radii larger than $0.5 \, r_{200}$, which is much beyond the radius of
our ISOCAM observations of A2218.

\section{Discussion}\label{s-disc}
\subsection{Cluster sources}\label{ss-disc-cl}
We have fitted the SEDs of 27 ISOCAM cluster sources (see
Table~\ref{t-sum}) with GRASIL models. Of these 27 sources, most are
characterized by a SED that declines from the NIR to the MIR, which is
a typical feature of quiescent galaxies with little or no star
formation. The SEDs of 20 ISOCAM cluster sources are best fit by
models without episodes of star formation in the last 1.1 Gyr.  For
simplicity, hereafter we refer to these 20 galaxies as the {\em
passive} cluster sample.

The MIR emission of these galaxies corresponds to the Rayleigh-Jeans
tail of the photospheric emission from cold stars, as found by Boselli
et al. (\cite{bose98}) for early-type galaxies in the Virgo
cluster. The MIR flux densities of the galaxies detected by Boselli et
al. (\cite{bose03}) in Virgo span a wide range ($\sim 0.1$--1000 mJy);
the 50\% completeness limits of our survey of A2218 at 6.7~$\mu$m
(see Paper~I) corresponds to 130~mJy at the distance of Virgo, hence
we have only detected the bright end of the distribution of early-type
galaxies in A2218 (see also Figs.~\ref{f-colcol} and \ref{f-cmr}). In
general, the optical morphologies of the {\em passive} cluster members
show they are early-type galaxies.

Our ISOCAM-selected {\em passive} galaxies seem to share the
properties of the average A2218 population of bright galaxies, for
which S01 estimated an age of 5--10 Gyr. An age of 5 Gyr implies that
the last major episode of star formation in these galaxies occurred at
$z \ga 0.9$.

The remaining seven ISOCAM-selected cluster sources show some
evidence of a (small) star formation activity. We refer to these
galaxies as the {\em active} cluster sample. Their SFRs are not
strong, $< 5$~$M_{\odot}$~yr$^{-1}$, and, according to the
best-fit GRASIL models, are significantly lower than the average SFRs
over the whole SFHs of these galaxies.  However, galaxies of the
{\em active} sample are among the faintest ISOCAM-detected members of
the A2218 cluster in the NIR (see Fig.~\ref{f-lirlnir}, top panel; the
{\em active} cluster sources are indicated with crosses). The
{\em active} cluster sources have IR luminosities that are comparable
to, or even higher than, their luminosities in the NIR.  Our results
are therefore consistent with the general trend found by Boselli et
al. (\cite{bose03}) in Virgo, namely that less massive cluster
galaxies have higher $F_{14.3}/F_{6.7}$ flux density ratios.

Morphologically, most of the {\em active} cluster galaxies are {\em
not} early-type, so it is perhaps not surprising that we find residual
traces of recent star formation activity in their SEDs. However, two
of them are S0--SB0/a. This is also not surprising, given the results
obtained by S01, who showed that 30\% of the S0 galaxies in A2218
have relatively blue colours, indicative of a recent (or ongoing) star
formation activity.  This supports a scenario in which part of cluster
S0s are formed via transformation from field spirals that have
recently entered the cluster environment (see, e.g., Poggianti
\cite{pogg03}).  We must caution the reader that four galaxies of the
{\em active} cluster sample are not spectroscopically confirmed
members, and it is therefore possible that some of them are in fact
field galaxies that we erroneously assign to the cluster (for one of
them, source 5, even the assigned optical counterpart is not certain,
see \S~\ref{s-data}).

The spatial and velocity distributions of the MIR cluster sources are
not different from those of the overall A2218 galaxy population. There
is no evidence that they belong to substructures of A2218 or to an
infalling population. Rather, it seems that the MIR selection of A2218
cluster sources is not biased. Probably, this happens because our MIR
cluster sample is dominated by the {\em passive} population selected
at 6.7~$\mu$m, where the emission is of stellar photospheric
origin. We would have obtained a different view of the cluster MIR
population had our 6.7~$\mu$m observations been less deep. If we only
consider the nine MIR cluster members detected at 14.3~$\mu$m, we find
that only two of them belong to the {\em passive} population.  It is
therefore important, when comparing A2218 with other clusters, to
account for the surveys different sensitivities and different areas.

In this respect, it is particularly interesting to compare the number
of cluster LIRGs detected by ISOCAM in five different clusters, namely
A1689 (Fadda et al. \cite{fadd00}; Duc et al. \cite{duc02}),
Cl0024+1654 (C04), A370 and A2390 (Paper~I), and A2218. The advantage
of using LIRGs to compare the number of IR active galaxies in
different clusters is that a LIRG luminosity is high enough to lie
above the limiting sensitivities of the surveys of all these
clusters. In C04 we have recently determined the number of LIRGs for
these five clusters, normalised by the surveyed areas and cluster
masses. The fraction of LIRGs in Cl0024+1654 turns out to be
significantly higher than the fraction in A370, and than the average
fraction in A1689, A2218, and A2390. There is however no significant
difference in the LIRG fractions of the latter three clusters.

In this same context it is also instructive to analyse the BO-effect
in A2218, compared to the BO-effect in Cl0024+1654 and A1689. In
Cl0024+1654 two thirds of the MIR cluster sources are part of the blue
BO galaxies, and the MIR cluster sources contribute a third of all BO
galaxies. In A1689, all BO galaxies are detected in the MIR (see
Fig.~7 in Duc et al. \cite{duc02}).  In A2218, only one MIR cluster
source qualifies as a blue BO galaxy, out of four cluster members.

The reason for the different LIRG fractions and MIR BO-effects among
different clusters is not clear. Part (but not all) of the differences
could be due to the different surveyed areas. While the LIRG fractions
were computed after normalising for the different surveyed areas, such
normalisation might not be sufficient if the LIRG fraction
significantly increases with the distance from the cluster
centre. Such a dependence is suggested by the analysis of
Cl0024+1654 (see C04), but more data are needed to confirm it.
On the other hand, the radial dependence of the BO-effect is an
obvious consequence of the morphology-density relation (e.g. Dressler
\cite{dres80}), and larger physical areas have been covered by ISOCAM
observations in Cl0024+1654 and A1689, than in A2218. Certainly,
a larger physical area of A2218 needs to be covered in the IR, before
we can make a solid statement about the (lack of a) MIR
BO-effect. After all, even in the optical the BO-effect remains
undetected when only the cluster central regions are observed
(Ellingson et al. \cite{elli01}), as in the present study. However,
given that {\em all} BO galaxies are MIR emitters in A1689, the
fraction of MIR emitters among BO galaxies would not change if the
surveyed area was smaller, and so it would still remain substantially
higher than the corresponding fraction in A2218.

The different LIRG fraction is not simply related to the average
evolution of clusters with redshift, since A370 and Cl0024+1654 have
similar redshifts and yet have very different LIRG fractions. In C04
we suggest that an ongoing merger could be at the origin of the
starbursting activity in the Cl0024+1654 galaxies.  However, also
A2218 is undergoing a merger with a substructure (Girardi et
al. \cite{gira97}; Cannon et al. \cite{cann99}; Machacek et
al. \cite{mach02}). An ongoing cluster-subcluster merger can stimulate
starbursts in cluster galaxies through the tidal effects of a rapidly
changing gravitational field (see, e.g., Bekki \cite{bekk01}), but
only if these galaxies are gas-rich. Differences in the average
cluster populations {\em before the merger occurs} could explain the
different impact of a cluster-subcluster merger on the stimulation of
star-formation and of IR-emission in cluster sources. As a matter of
fact, the fraction of emission-line galaxies in Cl0024+1654 (40~\%
according to Czoske et al. \cite{czos02}), is much higher than in
A2218, and in rich clusters in general (Biviano et al. \cite{bivi97}),
and, correspondingly, the fraction of blue galaxies is quite low in
A2218, possibly even lower than the average fraction in nearby
clusters (Rakos et al. \cite{rako01}).

We might thus speculate that the enhancement in the fraction of MIR
active cluster members is not simply the consequence of an ongoing
cluster-subcluster collision, but also of the integrated history of
the cluster, from which it depends how many cluster galaxies still
retain significant amounts of gas to fuel the star formation activity.
However, a larger physical area coverage of galaxy clusters in
the IR is needed before we can make a conclusive statement about the
variation (if any) of the number of MIR galaxies among different
clusters.

\subsection{Field sources}\label{ss-disc-fd}
We have fit the SEDs of 14 field sources(see Table~\ref{t-sum}) with
 GRASIL models. These SEDs are very different from those of cluster
 sources, and are characterized by an increasing flux density with
 increasing wavelength from the UV--optical to the MIR. Models of
 star-forming galaxies without an AGN component provide satisfactory
 fits to the observed SEDs. Although we cannot exclude that an AGN
 component is also present, it is unlikely to be dominant for these 14
 sources.  The SEDs of three field sources (see
 Appendix~\ref{a-sources}) could not be fit with any of the models we
 considered. Possibly this is due to a problem of source confusion,
 or, possibly, the emission of these three sources is
 AGN-dominated. In the latter case, the fraction of MIR sources in our
 sample with an AGN-dominated emission would be $\sim 0.2$, comparable
 to the fractions reported in the literature (Fadda et
 al. \cite{fadd02}; Elbaz et al. \cite{elba02}).  Characterizing the
 MIR SED of AGNs is however beyond the scope of this paper.

Models providing acceptable fits to the 14 field galaxy SEDs have
significant recent and ongoing star formation. To be specific, we find
that the median SFR of these 14 field galaxies is 22 $M_{\odot}
\, \mbox{yr}^{-1}$, and that this corresponds to 1.1 the SFR averaged
over the galaxies entire SFHs, according to the best-fit GRASIL
models. According to the same models, the SFRs of these galaxies have
not changed much over the last $\sim 1$ Gyr, so that $\sim 30$\% of
the galaxies total baryonic mass was converted in stars in the last
$\sim 1$ Gyr before the observing epoch (i.e. since $z \sim 0.8$,
taking the median redshift of these sources, $z=0.6$, as
representative of the sample). Such an intense star forming activity
can occur in a continuous smooth fashion or in a single starburst.  In
the former case, we would identify our MIR-selected field galaxies
with massive ($\sim 10^{11} \, M_{\odot}$) spirals near the peak of
activity in their SFH. In the latter case, the starburst is likely to
have occurred some time before observation, as SB models do not in
general provide best-fits to the observed field galaxy SEDs, and, in
fact, PSB models are preferred (see Table~\ref{t-sed} and
Figure~\ref{f-fstar}, bottom panel). Note however that since we only
considered two SB and two PSB models, the evidence in favour of PSB
vs. SB models is only tentative, but a full statistical approach of
the kind described by Chary \& Elbaz (\cite{char01}) is not justified
because of the rather large errors on our (quite faint) MIR fluxes.

Among the two rather extreme SFHs described above, a more realistic
one could be that described by Elbaz et al. (\cite{elba04}), in which
the galaxies undergo a series of Luminous IR Phases (LIRPs). Such
LIRPs would be produced by minor starburst events triggered by
galaxy-galaxy interactions, but not necessarily major collisions or
major mergers. This scenario is supported by the observations of
Flores et al. (\cite{flor99}) who have found that the SEDs of many
ISOCAM galaxies selected in the Canada-France Redshift Survey are not
indicative of young dusty starbursts, but, rather, of normal spirals
with a significant population of A stars, suggesting previous (recent)
starbursting episodes. Similarly, Mann et al. (\cite{mann02})
have shown that the SEDs of several ISOCAM sources in the Hubble Deep
Field South are adequately fitted by models of normal spirals, 
rather than by starburst models.

The morphologies of the field ISOCAM galaxies in our sample span a
wide range, from E to Sdm and Irr, but late-types dominate.
Peculiarities and asymmetries are visible in a few cases, possibly
indicating past or ongoing interactions with other galaxies, which
could be partly responsible for an enhanced star formation activity
(Condon et al. \cite{cond82}). Previous studies have already suggested
that interactions might be quite common among ISOCAM-selected field
galaxies (Flores et al. \cite{flor99}; Elbaz \& Cesarsky
\cite{elba03}; Hammer et al. \cite{hamm04}).

The LIRG fraction among our MIR field sources ($\sim 60$\%), and their
average SFR (33 $M_{\odot} \, \mbox{yr}^{-1}$) are in good agreement
with the values found in other ISOCAM field surveys (e.g. Chary
\& Elbaz \cite{char01}; Elbaz et al. \cite{elba02}; Elbaz \& Cesarski
\cite{elba03}; Mann et al. \cite{mann02}). In this context it is
useful to warn the reader about the different nomenclature used here,
with respect to the one adopted by Chary \& Elbaz (\cite{char01}) and
Elbaz et al. (\cite{elba02}). We all use the terms LIRGs and ULIRGs to
indicate, respectively, galaxies with $10^{11} \leq L_{IR}/L_{\odot} <
10^{12}$ and $L_{IR}/L_{\odot} \geq 10^{12}$. However, Chary \& Elbaz
(\cite{char01}) and Elbaz et al. (\cite{elba02}) use the term
'starburst' to indicate galaxies with intermediate IR luminosity
($10^{10} \leq L_{IR}/L_{\odot} < 10^{11}$), while we use the same
term to indicate that the galaxy is undergoing an intense and short
phase of star formation, at a rate significantly higher than the
average SFR of the galaxy before the starburst event, independently
from the galaxy IR luminosity. Adopting Elbaz et al.'s (\cite{elba02})
definition, we would count 8 LIRGs and 6 'starbursts' in our sample of
14 field galaxies.

The redshift distribution of our field sources ($0.1 \la z \la
1.1$, with a median $z \simeq 0.6$) is also in agreement with the
values found in other ISOCAM field surveys (see Flores et
al. \cite{flor99}; Rigolopolou et al. \cite{rigo02}; Elbaz \& Cesarski
\cite{elba03}; Paper~I).

In conclusion, our analysis confirms and reinforces previous results,
that we can summarize by saying that, on average, ISOCAM-selected
field galaxies are disk galaxies at $z \sim 0.6$, with ongoing star
formation activity at a rate of several dozens $M_{\odot} \,
\mbox{yr}^{-1}$. It is possible that some (or even most) of these
galaxies have suffered one or more starburst events in the
recent past, but only a minority seem to be detected during the
peak of the (short-lived) starburst phase. The contribution of
an AGN to the IR emission of field galaxies cannot be excluded, but it
is unlikely to be dominant.

\section{Summary and conclusions}\label{s-conc}
Using ISOCAM observations of the A2218 cluster (described at length in
Paper~I), optical data from LPS, S01, and Z01, and $HST$ imaging, we
have investigated the properties of MIR-selected sources in the A2218
cluster field. Specifically, we have used a battery of GRASIL models
(S98) to search for the best-fit to the optical-to-MIR SEDs of the
non-stellar sources. Acceptable and well constrained fits have been
obtained for the SEDs of 41 sources, of which 27 are cluster members
(see Table~\ref{t-sum}).

The properties of the cluster members detected in the MIR can be
summarized as follows. Most of them are early-type galaxies with
small or negligible ongoing star-formation, and no significant
evidence of recent previous episodes of star-formation. These galaxies
are mostly detected at 6.7~$\mu$m only, where we see the
Rayleigh-Jeans tail of cold stellar photospheres. A quarter of
the MIR cluster galaxies show significant (albeit small)
star-formation activity. They are characterized by smaller baryonic
masses (as indicated by their NIR luminosities), on average, and
generally have spiral or irregular morphologies, but two of them are
of S0--S0a type.

The spatial and velocity distributions of the MIR-selected cluster
galaxies are not significantly different from those of the general
cluster population. MIR-selected cluster galaxies are not, in general,
BO galaxies. Several independent analyses suggest that A2218 is not a
dynamically relaxed cluster. Complementing what is seen in other
clusters (see, e.g., C04), our results therefore suggest that a
cluster active dynamical status is not sufficient to affect the MIR
properties of its member galaxies.

Foreground and background galaxies in the A2218 cluster field are
mostly detected at 15~$\mu$m.  They span the redshift range 0.1--1.1,
with a median $z \simeq 0.6$.  Their MIR emission is most likely
dominated by dust-reprocessed stellar radiation, although we cannot
exclude an AGN contribution in some of them. Models of actively
star-forming galaxies, such as those of normal spirals and
starburst galaxies, observed close to the time of maximum
star-formation activity, but not necessarily during the short-lived
starburst phase, provide good fits to the SEDs of MIR field
galaxies. These field galaxies have a wide range of morphologies, but
most of them are spirals. Features suggestive of ongoing or past
interaction are found in a minority of the galaxies of the
sample. About half of the field galaxies in our sample are LIRGs, and
their high IR luminosities translate into a median SFR of $\sim 20 \,
M_{\odot} \, \mbox{yr}^{-1}$.

This paper is the second in a series devoted to the analysis of the
data of the ISOCAM gravitational lensing deep survey (see Paper~I).
In a forthcoming paper of these series (Perez et al. \cite{pere04}) we
will extend our analysis to the lensing clusters A370 and A2390,
which, together with our analyses of the clusters Cl0024+1654 (C04)
and A2219 (Coia et al.  \cite{coia04b}), will provide the most
extensive data-sample of MIR-selected galaxies in medium-distant
clusters. This sample will be expanded by forthcoming observations
with the Spitzer Space Telescope.

\begin{acknowledgements}
We thank Laura Silva for her invaluable advice on the use of the
GRASIL code, and Andy Pollock and Ian Smail for useful discussions. We
also wish to thank Jean Fran\c cois Le Borgne and Ian Smail for
sending us electronic versions of their data-sets. Finally, we thank
the referee for valuable suggestions that helped to improve the
presentation of our results.

The ISOCAM data presented in this paper was analysed using ``CIA'', a
joint development by the ESA Astrophysics Division and the ISOCAM
Consortium. The ISOCAM Consortium is led by the ISOCAM PI,
C. Cesarsky. This research has made use of NASA's Astrophysics Data
System, of the SIMBAD database at the Centre de Donn\'ees
Astronomiques in Strasbourg, and of the NASA/IPAC Extragalactic
Database (NED) which is operated by the Jet Propulsion Laboratory,
California Institute of Technology, under contract with the National
Aeronautics and Space Administration.
\end{acknowledgements}

\vspace*{1.0cm}

\appendix

\section{The SED fitting procedure}\label{a-sedfit}
As described in \S~\ref{s-sed}, we fit the SEDs of our ISOCAM sources
with 30 GRASIL models using a $\chi^2$ method. Here we
detail some aspects of the SED fitting procedure.

The wavelength resolution of the GRASIL models in the IR is such that
we must take the band-widths of the LW2 and LW3 ISOCAM filters (which
are significantly larger than the band-widths of optical and NIR
filters) into account, yet, on the other hand the detailed shape of
the filter wavelength responsivities (see, e.g., Blommaert et
al. \cite{blom01}) is not important, hence we consider them to be
box-shaped.

The model SEDs are redshifted to the source $z$. When the source $z$
is not known, we search for the best-fit among 12 values of $z$
between $z=0$ and $z=3$, roughly equally spaced in cosmological time,
and including the mean cluster redshift, $z=0.175$. Note that, at
variance with Mann et al. (\cite{mann02}), we only consider models
with an age less than the age of the Universe (in the adopted
cosmology) at the given $z$.

When the redshift of the optical counterpart is known, there are two
free parameters in the fit, the model SED and the normalisation of the
model. In principle, parameters in the GRASIL code depend on the
galaxy mass (e.g. the duration of the wind phase), so that the flux
scale cannot be re-normalized at will. However, a fine-tuning of the
GRASIL parameters is only appropriate when the galaxy SED we are
trying to fit is very well constrained. Otherwise, it is sensible to
leave the flux scale normalisation as a free parameter in the fit, in
so far as the observed galaxy masses are not very different from those
the models were originally designed to fit. The GRASIL models we use
here had their (mass-related) parameters set to reproduce galaxies of
$L^{\star}$ luminosity or brighter (L. Silva, private comm.), hence
they are appropriate for a comparison with our ISOCAM galaxies.

When $z$ is {\em not} known, it is taken as an additional free
parameter in the fitting procedure. In order to check how well we can
determine the photometric redshift, $z_{phot}$, of our ISOCAM galaxies
using the GRASIL models and the available optical-to-MIR data, we used
the ISOCAM sources with known-$z$, and left $z$ as a free parameter in
the SED-fitting procedure. The best-fit then provides our $z_{phot}$
estimate. The photometric and spectroscopic $z$'s are significantly
correlated (Kendall's rank correlation probability is $>99$~\%), and
the root mean square (rms, hereafter) difference between the
spectroscopic and photometric $z$'s is 0.10. Hence, we can hope to use
the SED-fitting procedure to determine approximate $z_{phot}$
estimates for the ISOCAM sources with the best-sampled SEDs.

In particular, for a given ISOCAM source we consider our $z_{phot}$
estimate sufficiently well constrained when the rms of the $z$-values
obtained in all the acceptable fits (at the $\geq 5$~\% c.l.) is
$z_{rms} \leq 0.2$. In this case, if $z_{phot} \pm z_{rms}$ is
consistent with the cluster mean redshift, we assign the source to the
cluster, otherwise, we adopt the best-fit $z_{phot}$ estimate.

Source confusion is not a major issue for most of our A2218
ISOCAM sources. However, when we suspect that another optical galaxy
could contribute to the MIR flux, we should consider the summed
contribution of both optical counterparts. Of course, in this case, we
must fit two models (and sometimes two redshifts) instead of
one. Hence, the number of free parameters sometimes become too large
with respect to the number of available data, and we cannot constrain
the best-fit solution. The available data suggest that we try the
two-model fit for source nos. 33, 45, 61a, 61b, 67.  In none of these
cases does the best-fit improve with respect to the fit with a single
optical counterpart.

\section{Notes on individual sources}\label{a-sources}
We discuss in some detail here the five sources whose SEDs
we failed to fit with the 30 GRASIL models considered (see
\S~\ref{s-sed}). 

{\bf ISO\_A2218\_02.} This ISOCAM source is detected in both the LW2
and LW3 bands. We identify this ISOCAM source with the optical
counterpart L664, at a redshift $z=0.53$, but we note that the nearby
galaxy L663, at unknown redshift, could contribute substantially to
the MIR flux, since its $r$-band flux density is 30\% of the flux density
of L664. Using L664 as the only optical counterpart, we do not find
acceptable fits to the source optical-to-MIR SED. Formally, the
best-fit is obtained for a Sa5 model, but the fit is poor mostly
because the model underestimates the $B$-band flux. It is possible
that the MIR flux is provided by both L664 and L663. However, we do
not have enough data to constrain a fit with two independent models
for the two galaxies, and a free redshift for L663.

{\bf ISO\_A2218\_28.} This ISOCAM source is detected in the LW2 band
only, and its optical counterpart is L391, the cluster cD, at
$z=0.1720$. Its spectral type is that of an early-type galaxy,
according to LPS.  The MIR emission of this ISOCAM source is very
extended, and probably it is not contributed solely by the cD (see
Fig.~\ref{f-imag1}). Using L391 as the only optical counterpart, this
source SED is best fit by an Em10 model, but the fit is poor, mainly
because it underestimates the $K_s$ band flux.  Given the very
extended emission from this galaxy, and the considerable crowding of
galaxies in its vicinity, it is perhaps not surprising that it is
difficult to fit its SED. Consistent aperture photometry is needed to
ensure that we are sampling the same parts of this galaxy at different
wavelength.  An accurate modelling of the cD is however beyond the
scope of this paper.

{\bf ISO\_A2218\_35.} This is an ISOCAM source detected in both the
LW2 and LW3 bands. It is identified with L317 (also S420), an Scd at
$z=0.474$. The best-fit to the source SED is given by a Mps5 model,
but the fit is only marginally acceptable (at the 1.6\% c.l.). 

{\bf ISO\_A2218\_53.} This ISOCAM source is detected in the LW3 band
only. Its most likely optical counterpart is L205 (also S368), but
L212 might contribute significantly to the MIR flux, since the
$r$-band flux ratio of the two galaxies is 0.4. L205 is an Sc at
$z=0.693$, L212 is an unknown-$z$ spiral. Using L205 as the only
optical counterpart, the best-fit is obtained for a Sc5 model, but the
fit is poor, mainly because it underestimates the $K_s$ band flux.
Unfortunately, we do not have enough data to constrain a fit with two
independent models for the two galaxies, and a free redshift for L212.
 
{\bf ISO\_A2218\_67.} This ISOCAM source is detected in both the LW2
and LW3 bands. It is identified with L75 (also S4004), a faint object
at unknown $z$. Other galaxies could contribute to the MIR emission,
in particular L71 (also S4005), an Scm, but possibly also L67. The
$r$-band flux ratios of L75 to L71 and to L67 are 1.4 and 0.6,
respectively. Adopting L75 as the only optical counterpart, we do not
find an acceptable fit to the source SED, whatever the source
redshift. Formally, the best-fit is found for an Es2 model at $z=1.5$,
but it is a poor fit that severely underestimates the MIR fluxes. We
try adding L71 as an additional counterpart.  Unfortunately, no
acceptable fits are found in this case either. The formal best-fit is
obtained for a combination of a $z=1.1$ El5 model, and a $z=2.8$ Es2
model. Although the fit now reproduces the LW2 flux, it still
underestimates the LW3 flux. The role of another possible counterpart,
L67, cannot be evaluated for lack of photometric data.  Notably, this
source has also been detected in the X-ray by Chandra (Sanchez,
private comm.). Hence, part of the MIR emission could arise from an
AGN component embedded in a dusty torus, something that GRASIL models
do not currently account for.

\end{document}